\documentclass[10pt,preprint]{article}
\usepackage{emulateapj,aastexug,natbib,psfig,rotating}

\newcommand{\etal} {et~al.}
\def\spose#1{\hbox to 0pt{#1\hss}}
\newcommand\lsim{\mathrel{\spose{\lower 3pt\hbox{$\mathchar"218$}}
     \raise 2.0pt\hbox{$\mathchar"13C$}}}
\newcommand\gsim{\mathrel{\spose{\lower 3pt\hbox{$\mathchar"218$}}
     \raise 2.0pt\hbox{$\mathchar"13E$}}}
\newcommand{\agn}{{\small AGN}}

\newcommand{\asca}{{\small \it ASCA}}

\newcommand{\ccd}{{\small CCD}}
\newcommand{\ce}{{\small CIE}}

\newcommand{\chandra}{{\it Chandra}}

\newcommand{\epic}{{\small EPIC}}
\newcommand{\esa}{{\small ESA}}

\newcommand{\gsfc}{{\small GSFC}}
\newcommand{\gsrp}{{\small GSRP}}

\newcommand{\heasarc}{{\small HEASARC}}

\newcommand{\hetgs}{{\small HETGS}}

\newcommand{\mrk}{{Markarian~3}}
\newcommand{\mos}{{\small MOS}}
\newcommand{\nasa}{{\small NASA}}

\newcommand{\ngc}{{\small NGC}~1068}

\newcommand{\nsf}{{\small NSF}}
\newcommand{\nwo}{{\small NWO}}

\newcommand{\pn}{{\small PN}}

\newcommand{\rec}{{\small REC}}
\newcommand{\rgs}{{\small RGS}}

\newcommand{\rrc}{{\small RRC}}
\newcommand{\sas}{{\small SAS}}
\newcommand{\sron}{{\small SRON}}
\newcommand{\usa}{{\small USA}}
\newcommand{\uv}{{\small UV}}
\newcommand{\xmm}{{\small \it XMM-Newton}}
\newcommand{\xspec}{{\small XSPEC}}
\newcommand{\xstar}{{\small XSTAR}}

\received{2002 March 18}
\begin{document}

\title{{\emph{XMM-Newton}} Reflection Grating Spectrometer Observations of 
Discrete Soft-X-ray Emission Features from NGC 1068}

\author{Ali Kinkhabwala\altaffilmark{1}, Masao Sako\altaffilmark{1,2,3}, Ehud 
Behar\altaffilmark{1}, Steven M. Kahn\altaffilmark{1}, Frits 
Paerels\altaffilmark{1}, Albert C. Brinkman\altaffilmark{4}, Jelle S. 
Kaastra\altaffilmark{4}, Ming~Feng Gu\altaffilmark{3,5}, Duane A. 
Liedahl\altaffilmark{6}}

\altaffiltext{1}{Columbia Astrophysics Laboratory, 
	      Columbia University, 
	      550 West 120th Street, 
	      New York, NY 10027; 
	      ali/masao/behar/skahn/frits@astro.columbia.edu}
\altaffiltext{2}{Present Address: Theoretical Astrophysics and Space Radiation Laboratory,
		California Institute of Technology,
		MC 130-33,
		Pasadena, CA 91125;
		masao@tapir.caltech.edu}
\altaffiltext{3}{Chandra fellow}
\altaffiltext{4}{Space Research Organization of the Netherlands, 
              Sorbonnelaan 2, 3548 CA, 
              Utrecht, The Netherlands; 
              a.c.brinkman/j.s.kaastra@sron.nl}
\altaffiltext{5}{Center for Space Research,
	      	Massechusetts Institute of Technology,
		Cambridge, MA, 02139;
		mfgu@space.mit.edu}
\altaffiltext{6}{Physics Department, 
              Lawrence Livermore National Laboratory,
              P.O. Box 808, L-41, 
              Livermore, CA 94550;
              duane@leo.llnl.gov}

\revised{}
\accepted{}

\begin{abstract}

We present the first high-resolution, soft-X-ray spectrum of the prototypical 
Seyfert~2 galaxy, \ngc.  This spectrum was obtained with the \xmm\ Reflection 
Grating Spectrometer (\rgs).  Emission lines from H-like and He-like low-Z 
ions (from C to Si) and Fe~L-shell ions dominate the spectrum.  Strong, narrow 
radiative recombination continua (\rrc) for several ions are also present, 
implying that most of the observed soft-X-ray emission arises in 
low-temperature plasma ($kT_{\mathrm{e}}\sim$ few~eV).  This plasma is 
photoionized by the inferred nuclear continuum (obscured along our line of 
sight), as expected in the unified model of active galactic nuclei (\agn).  
We find excess emission (compared with pure recombination) in all resonance 
lines (1s$\rightarrow$np) up to the photoelectric edge, demonstrating the 
importance of photoexcitation as well.  We introduce a simple model of 
a cone of plasma irradiated by the nuclear continuum; the line emission we 
observe along our line of sight perpendicular to the cone is produced through 
recombination/radiative cascade following photoionization and radiative 
decay following photoexcitation.  A remarkably good fit is obtained to 
the H-like and He-like ionic line series, with inferred radial ionic column 
densities consistent with recent observations of warm absorbers in Seyfert~1 
galaxies.  Previous \chandra\ imaging revealed a large (extending out to 
$\sim500$~pc) ionization cone containing most of the X-ray flux, implying 
that the warm absorber in \ngc\ is a large-scale outflow.  To explain the 
ionic column densities, a broad, flat distribution in the logarithm of the 
ionization parameter ($\xi=L_X/n_{\mathrm{e}}r^2$) is necessary, spanning 
$\log\xi=0$--3. This suggests either radially-stratified ionization zones, the 
existence of a broad density distribution (spanning a few orders of magnitude) 
at each radius, or some combination of both.  
\end{abstract}


\section{Introduction}

In the unified picture of active galactic nuclei (\agn) 
(Miller \& Antonucci 1983; Antonucci \& Miller 1985; Antonucci 1993), 
the soft-X-ray spectra of Seyfert~2 galaxies are expected to be
affected by emission and scattering from a medium strongly influenced by the 
nuclear continuum.  However, soft X-rays can also be produced 
through mechanical heating, as in shocks driven by supernova explosions in 
nuclear star-forming regions, or shocks created in or by outflowing material.  
It is quite plausible that both effects are important, i.e., the spectra
contain a combination of separate components of warm plasma whose 
emission is predominantly ``radiation-driven'' (dominated by photoionization 
and photoexcitation) and hot plasma whose 
emission is predominantly ``collision-driven'' (dominated by 
electron-impact ionizations and excitations at a temperature 
consistent with the observed line energies).   Since the same ionic 
transitions are expected in both types of plasma, distinguishing between 
these different emission mechanisms with the limited \ccd-type spectral 
resolution available on previous observatories has not been possible.

A recent, high-resolution \chandra\ \hetgs\ spectrum of the Seyfert~2, 
Markarian~3, showed unambiguously that much of the soft-X-ray emission from 
that source is produced through radiative recombination in photoionized plasma 
\cite{sako}.  Several resonance transitions (1s$\rightarrow$np), though, were 
stronger than expected for pure recombination.  An additional, hot, 
collisionally-ionized component was suggested by this excess.  However, the 
lack of accompanying strong Fe~L-shell lines argued against that possibility,
and was instead consistent with recombination in 
a photoionized plasma \cite{kallmanetal}.
Sako \etal\ (2000b) pointed out that the same nuclear continuum responsible 
for photoionizing the plasma may also photoexcite the ionic resonance 
transitions, producing line emission through radiative decay back to the 
ground state \cite{krolikkriss}.  Therefore, they asserted 
that the soft-X-ray emission from Markarian~3 could be explained in its 
entirety as reemission in warm plasma both photoionized and photoexcited by 
the inferred nuclear continuum.  This model is consistent with reprocessed 
emission from the warm absorbing medium typically observed in Seyfert~1 
galaxies, but now viewed from a different angle.

In this paper, we present the first high-resolution X-ray spectrum of the 
prototypical Seyfert~2 galaxy \ngc.
\ngc\ has a rich observational history, starting with its original, though less
commonly used, designation as M77 \cite{messier}.  Various independent 
studies in the optical early on confirmed its ``nebular''-type emission-line 
spectrum 
\cite{fath,slipher,hubble}.  It appeared as the first object in Seyfert's 
historic list \cite{seyfert} and, more recently, provided the first 
convincing evidence for a unified model of \agn\ \cite{antonuccimiller}.  
Extensive studies at all wavelengths have upheld this model:  
broad optical emission lines in polarized light 
\cite{antonuccimiller}, the existence of highly collimated 
bipolar radio jets (Wilson \& Ulvestad 1982; Wilson \& Ulvestad 1983; 
Pedlar et~al. 1989), biconical \ion{O}{3}\ emission regions \cite{ebstein}, 
reflected X-ray emission of an obscured nucleus (Monier \& Halpern 1987; 
Elvis \& Lawrence 1988; Koyama et~al. 1989), and a soft-X-ray component that 
lies significantly above the reflected continuum (Marshall et~al. 1993; Ueno 
et~al. 1994).

We show below --- using a new, simple radiative transfer code  --- that the
soft-X-ray spectrum of \ngc, obtained with the \xmm\ Reflection Grating 
Spectrometer (\rgs), can be explained in its entirety as 
emission from warm, irradiated plasma (as in Markarian~3), which 
reprocesses the nuclear continuum incident on it through 
recombination/radiative cascade following photoionization and radiative decay 
following photoexcitation.   

The format of our paper is as follows.  In \S\ref{sec:obs} we begin with 
a discussion of the observations and data analysis.    Detailed spectral 
analysis is presented in \S\ref{sec:spec}, followed by interpretation in 
terms of possible emission mechanisms in \S\ref{sec:int}.  In 
\S\ref{sec:novel}, we discuss a novel method for inferring 
geometrical/dynamical properties of the ionization cone from the observed 
relative amount of 
photoionization versus photoexcitation.  We discuss the remarkably 
consistent fits we obtain using this method in \S\ref{sec:finalfit}.  
Finally, in \S\ref{sec:dis}, we discuss the implications of these fits, 
including the connection between the observed large-scale ionization cone 
and generic warm absorbers.

\section{Observations and Data Reduction}\label{sec:obs}

\ngc\ was observed with \xmm\ for 110 ks starting at 5:48:05 
{\small UT} on 2000 July 29.  Data were taken simultaneously for the \epic\ 
detectors \mos\ and \pn, the optical monitor, and the \rgs.  All data were 
processed with the \xmm\ Science Analysis Software, \sas~5.2, with the
corresponding calibration files available for that version.  We concentrate in 
this paper mainly on the data from the \rgs, which was operated in its normal 
spectroscopy mode.  However, in order to confirm the 
negligible effect of source size and position on the extracted \rgs\ spectrum,
we also present a cursory analysis of the \pn\ image below.

The \rgs\ covers the wavelength range of 
approximately 6 to 38~\AA\ ($E=0.35$--$2.5$~keV) with a resolution of 
0.05~\AA, and a peak effective area of about 140~cm$^2$ at 15~\AA.
The \sas\ filters the observed \rgs\ events in dispersion channel versus 
\ccd-pulse-height space to separate the spectral orders.  The background is 
estimated using events from a region spatially offset from the source.  The 
wavelengths assigned to the dispersion channels are based on 
the pointing and geometry of the telescope with an overall accuracy of 
8~m\AA\ (1-$\sigma$) across the total wavelength range, which is
120~km~s$^{-1}$ at 20~\AA.  
Similarly, the estimated uncertainty in line broadening is 5~m\AA\
(1-$\sigma$), which is 75~km~s$^{-1}$ at 20~\AA.  Based on ground 
calibration, we expect the 1-$\sigma$
uncertainty in the effective area to be less than 10\% above 9~\AA\ and at 
most 20\% for shorter wavelengths \cite{herder,herder2}.

Independent, fluxed spectra from the \rgs~1 (red) and \rgs~2 (blue) 
instruments for all 110 ks of data for the $m=-1$ order are presented in 
Fig.~\ref{fig:spectrum}.  The spectral 
discontinuities are due to chip gaps, bad pixels, and the previous in-flight 
loss of one \ccd\ for \rgs~2 ($\lambda\sim20$--24~\AA).  Overall, the fluxed
spectra agree to $\lesssim10$\%.

\section{Spectral Analysis}\label{sec:spec}

\subsection{Line Emission}

\begin{figure}[]
\centerline{\psfig{figure=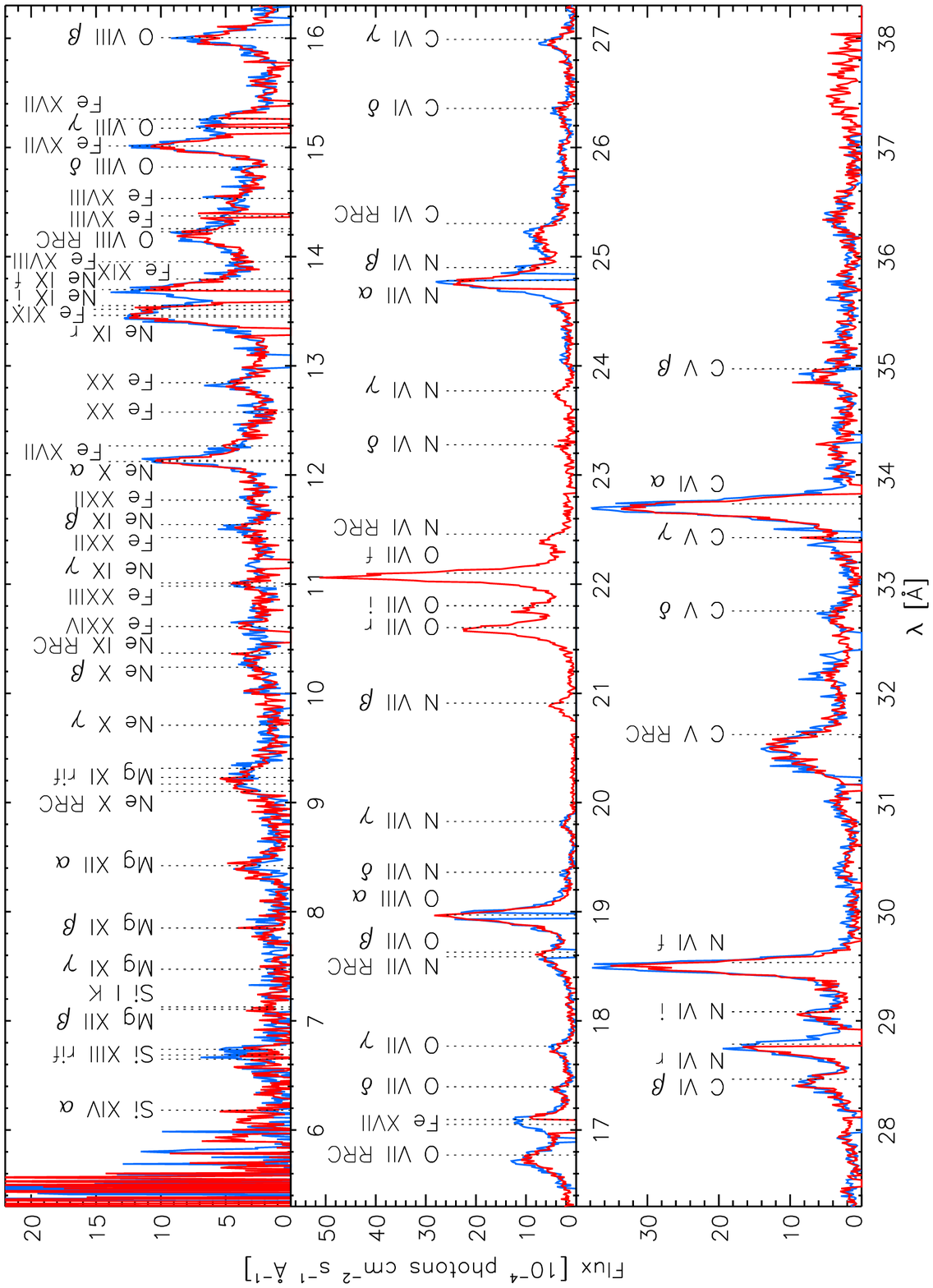,angle=180,width=6.5in}}
\caption{Effective-area-corrected, first-order RGS~1 (red) and RGS~2 (blue) 
spectra of NGC~1068 shifted to its rest frame ($z=0.00379$).  The spectral 
discontinuities are due to chip gaps in the CCD arrays, bad pixels, and the 
previous in-flight loss of one CCD for RGS~2 ($\lambda\sim20$--24~\AA).  Line 
labelling indicates the final state ion.  All H-like ($\alpha$) and He-like 
(r, i, and f) principal order lines are labelled for each ion with the 
corresponding RRC edges indicated as well.  Additionally, resonance 
transitions (np$\rightarrow$1s) are labelled as $\beta$ through $\delta$ 
(short for Ly$\beta$--$\delta$ and He$\beta$--$\delta$).  Several Fe~L-shell 
transitions are listed as well.  Unlabelled features at longer wavelengths 
(e.g., $\lambda=27.45,27.92, 30.4,31.0,$ 34.0--34.6, and 36.38~\AA) are 
likely due to L-shell tranisitions in mid-Z elements.  Line blueshifts are 
especially noticeable at longer wavelength.}
\label{fig:spectrum}
\end{figure}

The soft-X-ray spectrum of \ngc\ (Fig.~\ref{fig:spectrum}) is dominated by 
line emission.  Emission lines from H-like and He-like C, N, O, Ne, Mg, 
and Si are all clearly detected.   Numerous Fe~L-shell emission 
lines (\ion{Fe}{17} to \ion{Fe}{24}) are present as well.  Many higher-order 
resonance transitions (1s$\rightarrow$np) in H-like and He-like ions labelled 
Ly$\beta$--$\epsilon$ 
and He$\beta$--$\epsilon$, respectively, are prominent, with evidence for 
strong emission from even higher order transitions as well.  
Several unidentified features located in the bottom panel of 
Fig.~\ref{fig:spectrum} (e.g., at 27.92, 30.4, 34.0--34.6, and 36.38~\AA) are
likely due to L-shell emission from mid-Z ions such as Si or S.  A 
bright fluorescent line for Fe~K is observed in the \mos/\pn\ 
\ccd-resolution data, but the \rgs\ does not extend in wavelength down to 
Fe~K.  We report marginal detection of the fluorescent line of neutral Si in
the \rgs\ spectrum; there is no other evidence 
for any other significant fluorescent line emission.    We see no significant 
continuum emission in the spectrum.

\subsection{Line Table}

By performing joint fits to the separate $m=-1$ and $m=-2$ orders within 
\xspec\ \cite{arnaud}, we have measured line fluxes, widths, and shifts for 
all relatively bright lines which have been unambiguously identified 
(Table~\ref{tab:lines}).  
The measured fluxes take into account the Galactic column density of 
$N^{\mathrm{gal}}_{\mathrm H}=3.5\times10^{20}$~cm$^{-2}$ \cite{dickey} using
the neutral absorption model {\it tbabs} (Wilms, Allen, \& McCray 2000) in 
\xspec.  Line shifts are measured with 
respect to the systemic redshift of \ngc\ of $z=0.00379\pm0.00001$ or 
$cz=1137\pm3$~km~s$^{-1}$ (Huchra, Vogeley, \& Geller 1999).  

The brightness of \ngc\ coupled with the duration of the observation, implies 
that all the listed errors (representing $\pm$1-$\sigma$) of the 
unblended lines are dominated by instrumental uncertainty 
(see \S\ref{sec:obs}) and not by counting statistics.  For example, there are
approximately 6050 and 1880 total counts in the \ion{O}{7} forbidden line 
(\rgs~1 alone) and the \ion{Ne}{10} Ly$\alpha$ line (\rgs~1 and \rgs~2), 
respectively, implying counting errors of 1.3\% and 2.3\%; these errors are 
much less than the effective area uncertainty of 10\%.

\subsection{Line Shifts and Widths}

We find that most emission lines in the \rgs\ spectrum are shifted and appear 
broader than expected for monochromatic lines (Table~\ref{tab:lines}).

\begin{figure*}[]
\centerline{\psfig{figure=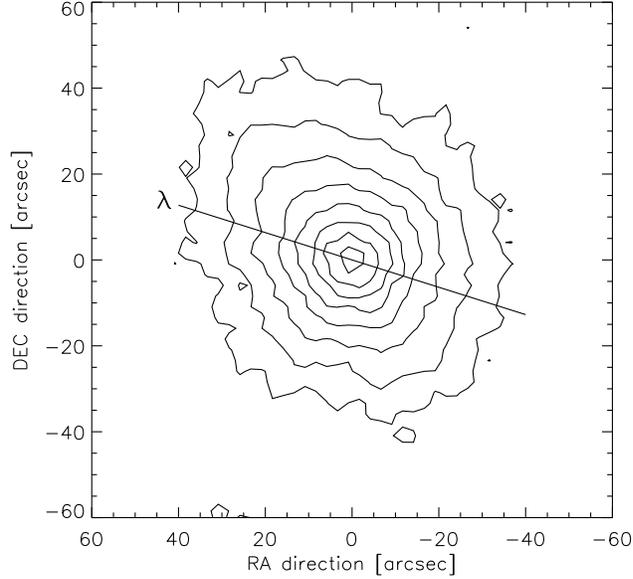,angle=90,width=4.5in}}
\caption{Image of NGC~1068 taken with the PN camera on board {\it XMM-Newton}. 
Only X-rays with CCD-determined energies less than 
2.5~keV were taken.
We use factor-of-two contours with the lowest at $2.8\times10^{-4}$ counts 
arsec$^{-2}$ s$^{-1}$. The image is centered at the peak of emission at
$\alpha$(J2000)=02$^{\mathrm{h}}$42$^{\mathrm{m}}$40$^{\mathrm{s}}$.65, 
$\delta$(J2000)=$-$00$^{\circ}$00\arcmin41\arcsec.40 ($1\arcsec=72$~pc).  
The RGS dispersion axis is indicated (oriented $72.34^\circ$ counterclockwise 
from the vertical), with $\lambda$ indicating the direction of increasing 
wavelength.}
\label{fig:contour}
\end{figure*}

The line shifts are due to velocity shifts at the source and not, for 
example, due to an absolute pointing error.  The position of the nominal 
centroid of the X-ray emission in the \pn\ image with a \ccd -determined 
energy cut consistent with the \rgs\ waveband (Fig.~\ref{fig:contour}) has
been used to derive the \rgs\ spectrum.  Gaussian fits to the \pn\ image 
using several different \ccd -resolution energy cuts 
yield centroids all within 0\arcsec.5 of each other, proving that possible 
spatial stratification of different emission zones is insignificant.
The overall positional uncertainty between
the \pn\ and \rgs\ instruments is less than the absolute pointing 
uncertainty of the telescope of 4\arcsec, implying a robust upper limit on the
line shift uncertainty of 8~m\AA\ (120~km~s$^{-1}$ at 20~\AA).

Significant velocity blueshifts from 
roughly $0$ to $600$~km~s$^{-1}$ have been measured from the \rgs\ spectrum 
(Table~\ref{tab:lines}). 
Line blueshifts of each ionic line series are fairly consistent.  
Longer wavelength lines appear to have higher blueshifts.  Overall, the
observed blueshifts are comparable to blueshifts of optical/\uv\ emission 
lines 
(Grimes, Kriss, \& Espey 1999; Kraemer \& Crenshaw 2000), though, in contrast 
to these observations, we find no evidence for any significant redshifts. 

\begin{figure*}[]
\centerline{\psfig{figure=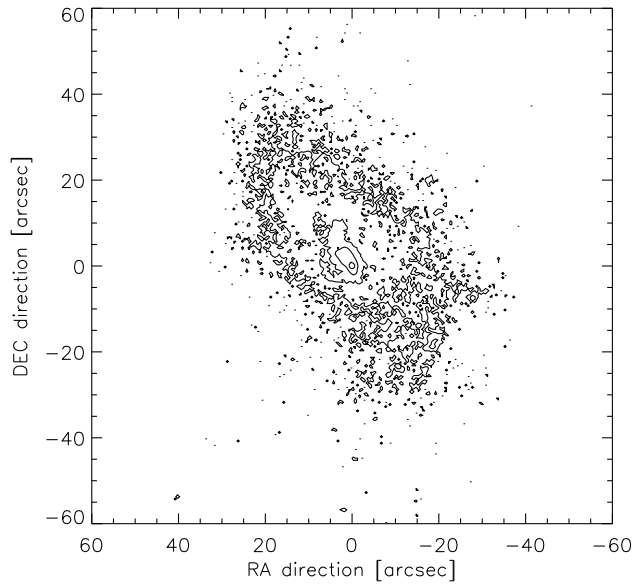,angle=90,width=4.5in}}
\caption{{\it Chandra} ACIS-S contour image of NGC~1068 using a frame time of 
0.4~s (Young \etal\ 2001).  Only X-rays with CCD-determined energies less than 
2.5~keV were taken.  We 
use factor-of-ten contours with the lowest at 
$3.5\times10^{-4}$~counts~arsec$^{-2}$~s$^{-1}$.  
The image is centered on the peak of the X-ray emission 
at $\alpha$(J2000)=02$^{\mathrm{h}}$42$^{\mathrm{m}}$40$^{\mathrm{s}}$.71, 
$\delta$(J2000)=$-$00$^{\circ}$00\arcmin47\arcsec.9 ($1\arcsec=72$~pc).  
\label{fig:contour_large}}
\end{figure*}

The widths we measure are similarly due to intrinsic velocity distributions at 
the source.  A \chandra\ image of \ngc\ (Young, Wilson, \& Shopbell 
2001) shows evidence for weak extended emission on scales of tens of 
arcseconds (Fig.~\ref{fig:contour_large}).
The effect of spatial broadening in the \rgs\ spectrum for this
source, characterized by a strongly peaked central component and weak extended 
emission, is negligible (A.~Rasmussen, private comm.).  Therefore, any 
excess broadening of particular lines can only be due to velocity 
distributions at the source.
Measured velocity widths are also presented in Table~\ref{tab:lines}.  
All widths lie in the range $\sigma_v^{\mathrm{obs}}=300$--$700$~km~s$^{-1}$, 
which is consistent with the linewidths associated with the 
narrow-line-emitting regions observed in the \uv\ 
(Grimes, Kriss, \& Espey 1999).

\subsection{Radiative Recombination Continua}

The spectrum also includes very distinctive radiative recombination 
continua (\rrc) for H-like and He-like C, N, and O, which
are produced when electrons recombine directly to the ground state 
of these highly-ionized species.  \rrc\ are broad features for hot,
collisionally-ionized plasma, but are narrow, prominent features for cooler 
photoionized plasma.  The narrow width of these \rrc\ provide a direct measure 
of the recombining electron temperature \cite{liedahlpaerels,liedahl}.

For those \rrc\ which are clearly detected (\ion{C}{5}, \ion{C}{6}, 
\ion{N}{6}, \ion{N}{7}, \ion{O}{7}, and \ion{O}{8}), we were able to determine 
accurate temperatures from the \rrc\ profile itself (Table~\ref{tab:rrc}).  
The unblended \rrc\ (on the short wavelength side) of \ion{C}{5} and 
\ion{O}{7} provide the best temperature determinations.  For the other \rrc, 
the dominant source of uncertainty comes 
from blending with other nearby lines.  All temperatures are consistent with 
the conservative range of $kT_{\mathrm{e}}\sim$ few~eV.  These temperatures 
imply velocity broadening of all features by $\sim10$~km~s$^{-1}$.  This 
broadening is much lower than the observed broadening of 
$\sim400$~km~s$^{-1}$, implying the importance of bulk and/or turbulent cloud 
velocities.  

Higher temperatures of $kT_{\mathrm{e}}=20$--30~eV were inferred from the 
broad, polarized, electron-scattered optical lines 
(Miller, Goodrich, \& Mathews 
1991).  But these temperatures are really only upper limits, since nonthermal 
broadening due to bulk and/or turbulent cloud velocities, as observed in the 
X-ray, could also be present.  

Weak features at the positions of the \rrc\ from \ion{Ne}{9} and \ion{Ne}{10} 
are also detected, though severe blending makes a temperature determination 
difficult in these cases.  Their inferred temperatures are not necessarily 
discrepant with the range given for the other ions, though significantly 
hotter temperatures cannot be excluded.

\subsection{Higher-order Series Lines}

In addition to the prominent $\mathrm{n}=2\rightarrow1$ transitions in the 
H-like and He-like ions, the unique combination of high resolution and large 
effective area characteristic of the \rgs\ allows for the detection and flux 
measurement of several discrete higher-order resonance transitions 
(np$\rightarrow$1s), as well as blended higher-order transitions leading 
up to the \rrc\ edge.  For the H-like ions, we list the observed ratios of 
these transitions to the Ly$\alpha$ line in Table~\ref{tab:Hratios}, where we 
also give the ratios expected for recombination in a warm, photoionized plasma and
for hotter plasma in collisional equilibrium.  For the He-like ions, we 
list the observed ratios of these transitions to the forbidden line in 
Table~\ref{tab:Heratios}, together with the ratios expected for 
recombination.  Here, we do not provide the ratios expected for a hotter 
plasma in collisional equilibrium, because the strong dependence of the 
relative strength of the forbidden line on temperature would make that 
comparison meaningless.

\subsection{He-like Triplet Ratios}

For He-like N and O, we present measurements of the standard plasma
diagnostic ratios $R=f/i$ and $G=(f+i)/r$ \cite{gabriel} in 
Table~\ref{tab:rif}, where $r$, $i$, 
and $f$ are the resonance, intercombination, and forbidden line fluxes.  
Also listed are the
ratios expected for recombination in a warm, photoionized plasma and for 
hotter plasma in collisional equilibrium.

\section{Interpretation of the Observed Spectral Features}\label{sec:int}

\begin{figure*}[]
\centerline{\psfig{figure=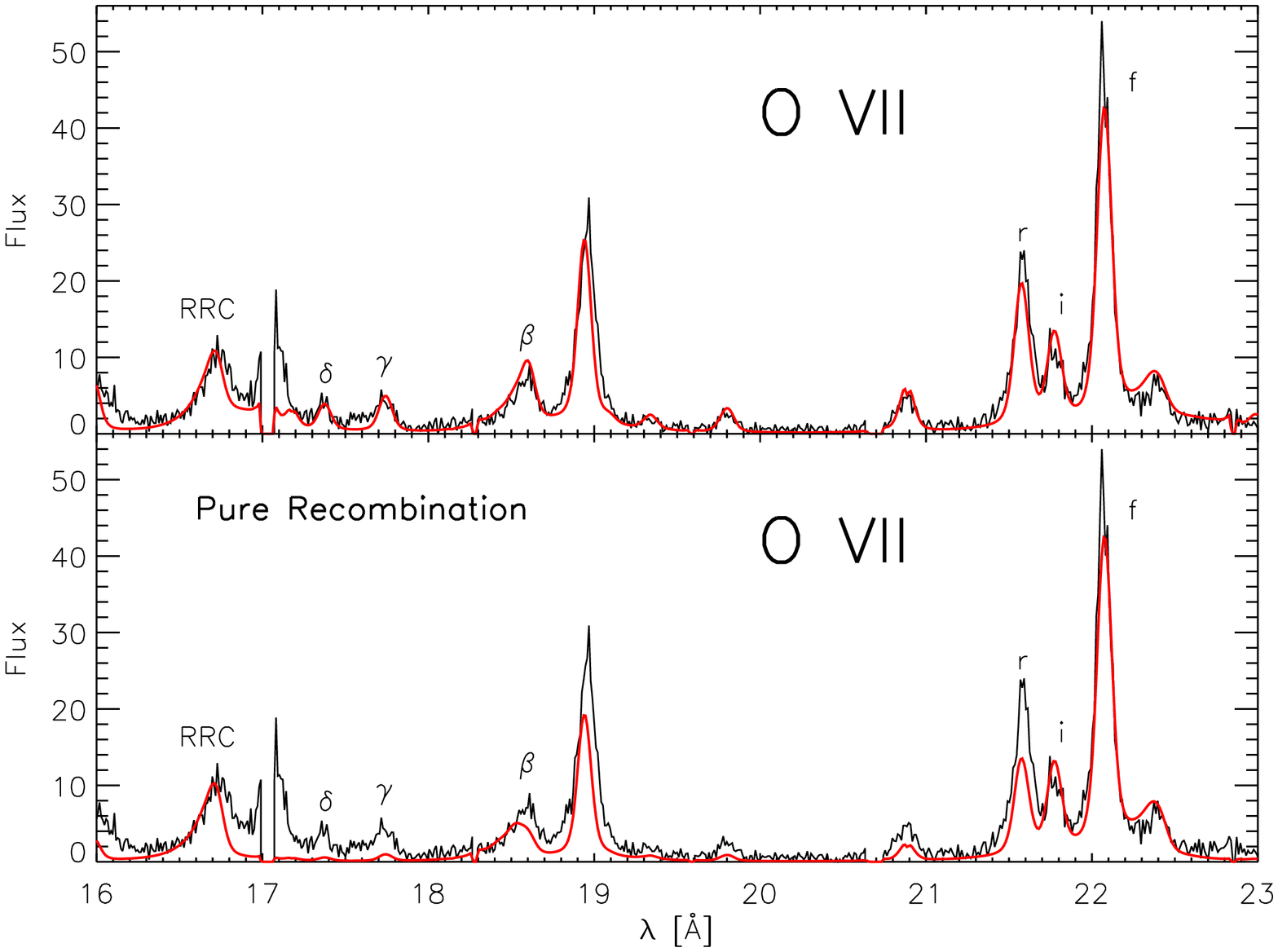,angle=90,width=3.9in}}
\caption{Final fit parameters for \ion{O}{7} (see Table~\ref{tab:finalfit} and 
Fig.~\ref{fig:finalfit}) including recombination/radiative cascade
following photoionization and radiative decay following photoexcitation 
(top).  Recombination alone (bottom) is unable to explain the excess 
emission in all resonance lines np$\rightarrow$1s.  (Data are plotted in rest 
frame of NGC~1068.  The model spectrum has been velocity shifted by 
$-400$~km~s$^{-1}$.)}
\label{fig:oxygen}
\end{figure*}

Unambiguous evidence for the predominance of emission in a photoionized 
plasma comes 
primarily from the observed bright \rrc, which indicate recombination at a 
temperature around a few eV.  In addition, as we explain in detail
below, the ratios of the intercombination and forbidden lines to the \rrc\ are 
very close to the values expected for pure recombination (see 
\S\ref{sec:oxygen} and bottom panel of Fig.~\ref{fig:oxygen}).  The $R$ ratio 
for \ion{N}{6} (Table~\ref{tab:rif}) 
also appears to favor pure recombination, though the $R$ ratio for \ion{O}{7} 
is somewhat ambiguous (see \S\ref{sec:oxygen}).

The anomalous He-like $G$-ratios (Table~\ref{tab:rif}), however, are 
grossly inconsistent 
with pure recombination, suggesting an additional component and/or mechanism.
An additional, weak collisionally-ionized plasma component, 
whose He-like triplets generically exhibit a higher resonance line 
intensity compared 
to the intercombination and forbidden lines, could explain these effects.  
However, such a component would not only enhance the 
resonance line, but {\em all} lines in the triplet, thereby
overpredicting the intercombination and forbidden line fluxes relative to the
\rrc.  This interpretation is marginally ruled out by the triplet to \rrc\ 
ratio in the \rgs\ data.

\begin{figure*}[]
    \centerline{\psfig{figure=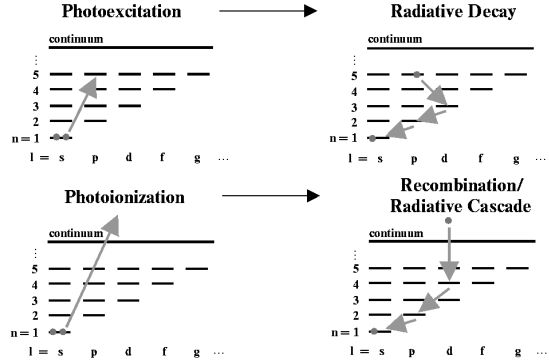,angle=0,width=3.5in}}
\caption{Atomic processes of photoexcitation and photoionization with inverse 
processes of radiative decay and recombination/radiative cascade, 
respectively.  Most radiative decays after photoexcitation will occur directly 
to the ground state, but other paths (as shown in the diagram) are also 
possible.}
\label{fig:grotrian}
\end{figure*}

Similarly ambiguous $G$ ratios were previously observed in the 
\chandra\ \hetgs\ spectrum of the Seyfert~2, \mrk. 
Sako \etal\ (2000b) argued against the presence of an additional 
collisionally-ionized 
plasma component due to the observed weakness of the Fe~L 
shell emission, which was consistent with recombination in a 
photoionized plasma \cite{kallmanetal}.  Instead, 
Sako \etal\ (2000b) suggested that the enhanced resonance lines 
were due to photoexcitation (see Fig.~\ref{fig:grotrian} for a schematic of
the relevant atomic processes).  This argument is sensitive to the 
Fe abundance; however, Sako \etal\ (2000b) noted that a rather unlikely 
order-of-magnitude underabundance of Fe would be necessary to be consistent
with the presence of additional hot collisionally-ionized plasma.

\begin{figure*}[]
    \centerline{\psfig{figure=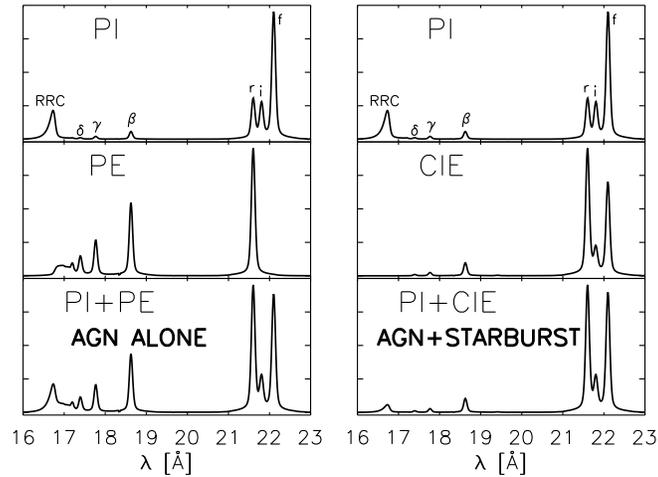,angle=90,width=3.5in}}
\caption{We demonstrate how to distinguish between hot collisional plasma 
(e.g., starburst region) and photoexcitation with the He-like O~VII line
series.  Starting with pure recombination following photoionization 
(``PI'' -- top two panels, $kT_{\mathrm{e}}=4$~eV), we self-consistently add 
radiative decay following photoexcitation (``PE''-- middle left panel),
assuming reasonable ionization cone parameters (as in \S\ref{sec:novel}), or 
an additional hot plasma component in collisional ionization equilibrium 
(``CIE'' -- middle right panel, $kT=150$~eV), for which we obtain the bottom 
two panels.  Note that both bottom panels have similar triplet ratios, 
implying that consideration of the triplet alone is insufficient to 
discriminate between the two scenarios.
However, the ``AGN ALONE'' panel has significantly stronger 
higher-order-series transitions (including the RRC) than the ``AGN+STARBURST'' 
panel, demonstrating the diagnostic importance of these transitions.  
Photoexcitation generically enhances higher-order-series transitions beyond 
the amount possible in a hot, collisionally-ionized plasma due to the 
flatness of the incident photoionizing spectrum (compared with a Maxwellian 
electron distribution) as well as the characteristic ``curve of growth''
effects obtained in irradiated plasmas (see Fig.~\ref{fig:opticaldepth}).
(Normalization in each panel is arbitrary.)}
\label{fig:PIPECIE}
\end{figure*}

For \ngc, we propose a cleaner test based solely on each particular ionic line
series.  The key to robustly distinguishing between photoexcitation and 
additional collisionally-ionized plasma components lies in the higher-order 
series line strengths.  We illustrate this new method in 
Fig.~\ref{fig:PIPECIE}.  
Starting with a pure recombination spectrum, we add in photoexcitation 
(self-consistently, as presented below in \S\ref{sec:novel}) or a separate 
collisionally-ionized plasma component.  The
addition of photoexcitation enhances all resonance transitions, whereas the
addition of collisionally-ionized plasma primarily enhances only the triplet
lines (with more weight on the resonance line of the triplet).

The ratios of higher-order-series lines to 
Ly$\alpha$ for \ion{C}{6}, 
\ion{N}{7}, and \ion{O}{8} shown in Table~\ref{tab:Hratios} are 
significantly higher than those expected for
collisional ionization equilibrium (\ce), though they are not entirely 
consistent with pure 
recombination (\rec) either.  Photoexcitation boosts all lines in this series 
including Ly$\alpha$, but its effect is dependent on the column density and 
velocity distribution, so it must be modeled carefully.

However, the forbidden line in He-like ions is unaffected by
photoexcitation, allowing for a measure of the significance
of photoexcitation of higher-order-series transitions with respect to
the forbidden line, which is fed through recombination after 
photoionization.  
Comparison of the ratios of these higher-order-series lines to the forbidden 
line (Table~\ref{tab:Heratios}) with similar ratios for plasma in collisional 
equilibrium is not as
useful due to the significant temperature sensitivity of $f$.  However, 
ratios for plasma in collisional equilibrium are generally comparable to or
lower than ratios for pure recombination (Fig.~\ref{fig:PIPECIE}).  
Therefore, the 
excess strength of the higher-order lines, especially the He$\gamma$ and 
He$\delta$, can only be explained by photoexcitation.

\section{Novel Diagnostics from Photoionization and Photoexcitation Rates}\label{sec:novel}

We now show that the relative amount of photoionization versus 
photoexcitation inferred from the soft-X-ray spectrum of \ngc\ provides 
detailed information about the geometry of absorbing/reemitting material
around the central, power-law-continuum source.

\begin{figure}[]
\centerline{\psfig{figure=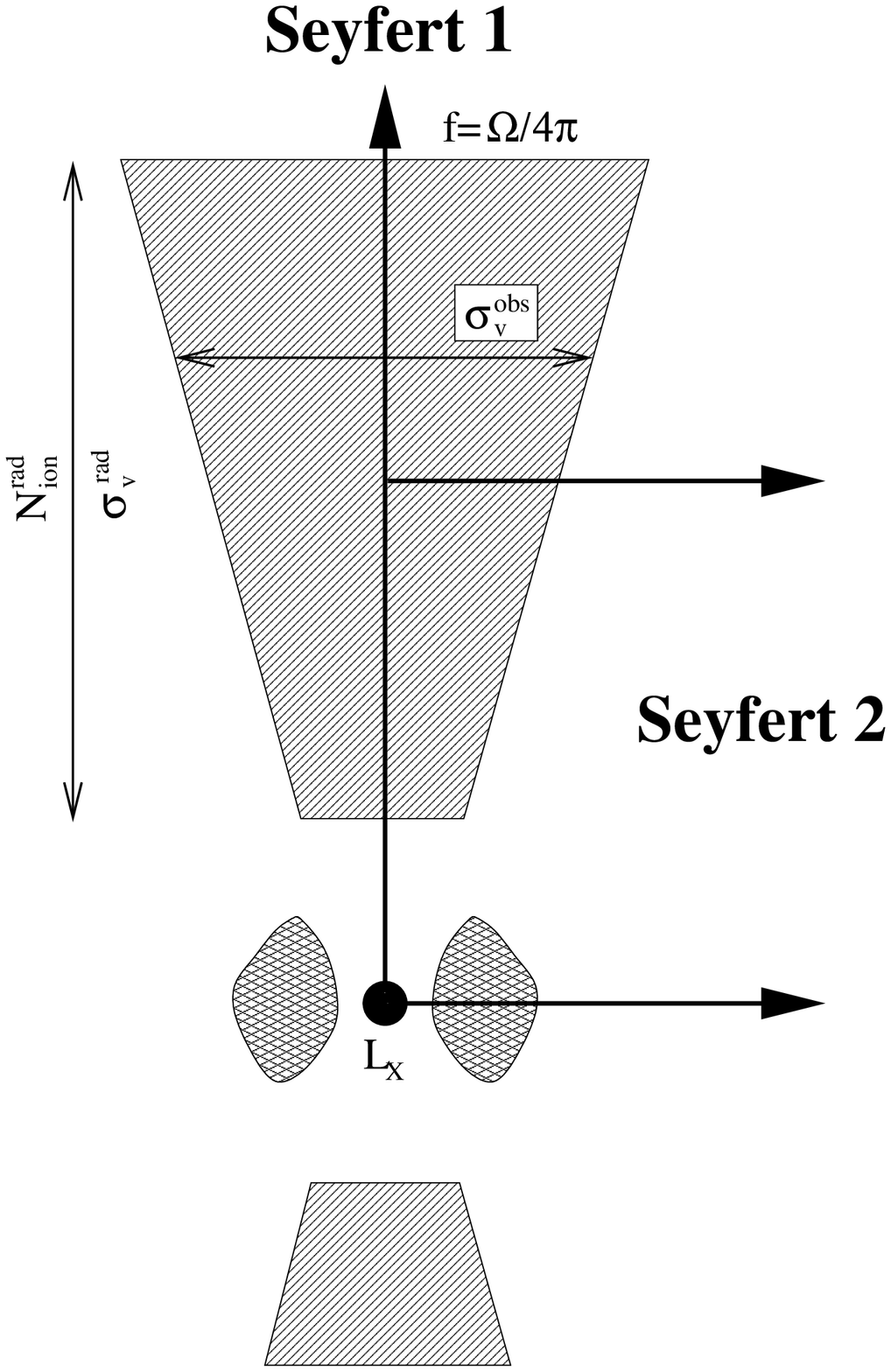,angle=0,width=5.in}}
\caption{Simple cartoon of ionized cone model (not to scale).  The central 
nuclear component, denoting the black-hole/accretion-disc/comptonized-halo 
system, is shown as the black spot.  In the Seyfert~1 view, the ionized
cone (warm absorber) is seen through its absorption of the intrinsic 
nuclear continuum $L_X$.  In the Seyfert~2 view, the intrinsic continuum is 
highly absorbed by intervening material (``dusty torus'' drawn to the right 
and left of the nucleus), allowing for the reprocessed emission in the 
ionized cone to be observed.  The variables comprising the two global 
parameters $\sigma_v^{\mathrm{rad}}$ and $fL_X$ in our model are indicated.
The radial ionic column density $N^{\mathrm{rad}}_{\mathrm{ion}}$, which is
left free for each ion, is also indicated.  The Seyfert-2-observed
velocity distribution perpendicular to the cone is $\sigma_v^{\mathrm{obs}}$.  
For NGC~1068, the intrinsic continuum is likely completely obscured, 
contributing no flux to the soft-X-ray regime.  Also, in NGC~1068, the NE 
cone is much brighter than its counterpart in the SW (hence the asymmetry in 
the above diagram), therefore, the covering factor $f$ applies to the NE cone 
alone.  Generic Seyfert~2 galaxies, however, may show evidence for two equally 
bright ionization cones.}
\label{fig:cone}
\end{figure}

In the unified model of active galaxies, the 
observable properties of a particular \agn\ depend primarily on orientation.
The soft-X-ray warm absorber observed in Seyfert~1 galaxies,
therefore, should show itself in reemission for the roughly perpendicular 
line of sight to Seyfert~2 galaxies (see Fig.~\ref{fig:cone}).
For the Seyfert~1 view, the relative 
amount of edge absorption versus line absorption 
depends greatly on the column density through the material.  
Thus, the Seyfert~2 spectrum should similarly be sensitive to the radial 
column density, since each photoionization is balanced by
recombination/radiative cascade, and each photoexcitation 
is similarly balanced by radiative decay.
Therefore, the shape of the reemitted Seyfert~2 ionic line series can be used 
to infer the radial (or Seyfert-1-observed) ionic column densities through the 
warm absorber.

For low radial ionic column density, $N^{\mathrm{rad}}_{\mathrm{ion}}$, all 
lines and photoelectric edges remain 
unsaturated.  For a H-like or He-like ion, the summed 
oscillator strength for transitions to discrete levels is 
roughly comparable to the integrated oscillator strength for transitions 
to the continuum (e.g., Bethe \& Salpeter 1977), implying
roughly similar amounts of radiative decay following photoexcitation and 
recombination/radiative cascade following photoionization (modulated by 
the shape of the continuum at the relevant photoexcitation/photoionization 
energies).  

For higher $N^{\mathrm{rad}}_{\mathrm{ion}}$, the low-lying, 
high-oscillator-strength transitions are the first to saturate, leading to 
a ``tilting'' of the photoexcitation contribution to the Seyfert~2 spectrum to 
lower-oscillator-strength, higher-order-series transitions.  This can
simply be thought of as the standard ``curve of growth,'' except now observed
indirectly through reemission.  Due to this gradual line saturation and the 
fact that the edge always saturates last, 
it is evident that an increase in $N^{\mathrm{rad}}_{\mathrm{ion}}$ implies an 
increase in the overall relative importance of photoionization compared to 
photoexcitation.

At very high $N^{\mathrm{rad}}_{\mathrm{ion}}$, all discrete transitions are 
fully saturated and the reemitted spectrum is that of pure recombination 
following photoionization.  In the following discussions, these qualitative 
arguments are made more explicit. 

\subsection{Radial Ionic Column Density}

\begin{figure}[]
\centerline{\psfig{figure=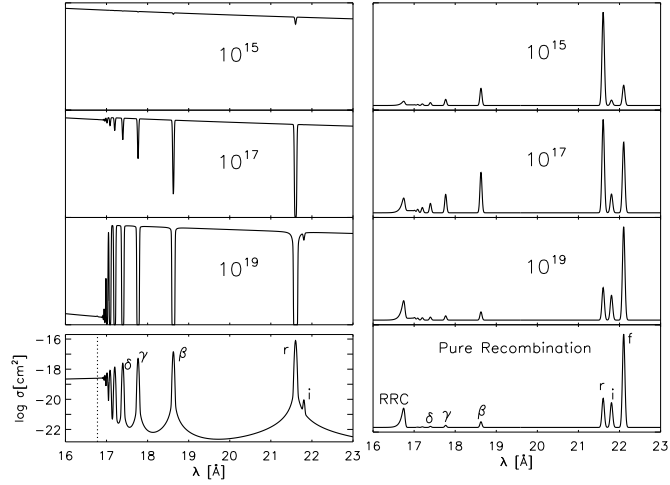,angle=90,width=3.5in}}
\caption{Effect of differing radial column densities on the reemitted 
spectrum for He-like O~VII.  The top three panels on the left show the radial 
Seyfert~1 view through the outflow and down to the nucleus for radial column 
densities in O~VII of $10^{15},10^{17},$ and $10^{19}$~cm$^{-2}$.  (The 
logarithm of the O~VII cross section for photoexcitation and photoionization 
with separating boundary is presented in the lower left panel.)  Corresponding 
panels on the right show the Seyfert~2 view roughly perpendicular to the axis 
of outflow and from which the nucleus is completely obscured 
(Fig.~\ref{fig:cone}).  All photons absorbed out of the power law in the 
Seyfert~1 spectrum are reprocessed and reemitted to generate the Seyfert~2 
spectrum.  Radiative decay following photoexcitation dominates the Seyfert~2 
spectrum at low column densities, whereas recombination/radiative cascade 
following photoionization dominates at high column densities (for comparison, 
pure recombination is shown in the lower right panel).  For H-like ions, the 
behavior is similar (omitting the intercombination/forbidden lines).  
Throughout we take a radial gaussian distribution with 
$\sigma^{\mathrm{rad}}_v=200$~km~s$^{-1}$, Seyfert-2-observed velocity 
distribution with $\sigma^{\mathrm{obs}}_v=400$~km~s$^{-1}$, and temperature 
$kT_{\mathrm{e}}=3$~eV.  (Normalization in each panel is arbitrary.)}
\label{fig:opticaldepth}
\end{figure}

The dependence of our model ionic spectra on
radial ionic column density, $N^{\mathrm{rad}}_{\mathrm{ion}}$, is 
illustrated in Fig.~\ref{fig:opticaldepth}. Here we show 
the effect of varying $N^{\mathrm{rad}}_{\mathrm{ion}}$ for He-like 
\ion{O}{7}.

The Seyfert~1 and Seyfert~2 views are shown in the top three panels on 
the left and the right of Fig.~\ref{fig:opticaldepth}, respectively, for 
different radial ionic column densities.
The total edge absorptions (photoionizations) and line absorptions 
(photoexcitations) in the Seyfert~1 view yield the expected amounts of 
recombination/radiative cascade following photoionization and 
radiative decay following photoexcitation in the Seyfert~2 view.
Note the relative strength of the lines to the \rrc\ and the effect of 
saturation on the high-oscillator-strength, lower-order-series transitions.  
In particular, 
note the strong variation in the relative strength of the resonance line in 
the He-like triplet as a function of optical depth, illustrating that the 
triplet, by itself, is a poor diagnostic for determining whether or not 
hotter, collision-driven plasma (which would show a strong resonance 
line) is present (see Fig.~\ref{fig:PIPECIE}).

\subsection{Radial Velocity Width}

\begin{figure*}[]
\centerline{\psfig{figure=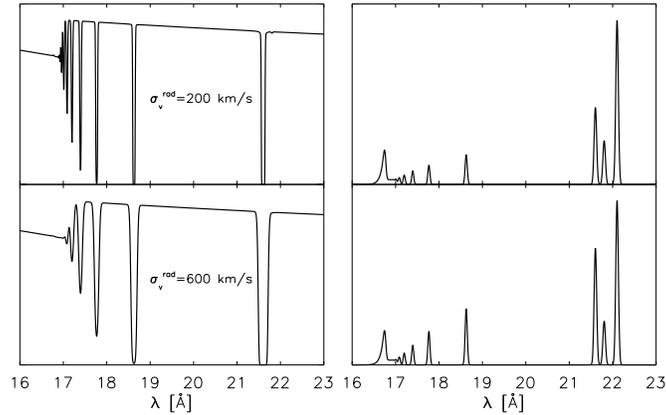,angle=90,width=3.7in}}
\caption{Effect of different $\sigma_v^{\mathrm{rad}}$ 
for He-like O~VII spectrum using a radial column density of 
$10^{18}$~cm$^{-2}$.  At larger $\sigma_v^{\mathrm{rad}}$, photoexcitation is 
enhanced relative to photoionization.  Spectra on the right-hand side were 
convolved with the same transverse velocity distribution 
($\sigma_v^{\mathrm{obs}}=400$~km~s$^{-1}$) and assume $kT_{\mathrm{e}}=3$~eV.}
\label{fig:sigmav}
\end{figure*}

The radial velocity width specified by $\sigma_v^{\mathrm{rad}}$ is a model
parameter.  There is little {\em a priori} reason to believe that 
$\sigma_v^{\mathrm{rad}}$ should be identical to the {\em observed} 
velocity width of the lines, $\sigma_v^{\mathrm{obs}}$, viewed perpendicular 
to the cone (see Fig.~\ref{fig:cone}).  For this
reason, we keep $\sigma_v^{\mathrm{rad}}$ a free parameter.  The 
effect of varying $\sigma_v^{\mathrm{rad}}$ is illustrated in 
Fig.~\ref{fig:sigmav}.  Increasing $\sigma_v^{\mathrm{rad}}$ 
by a certain factor leads to a roughly similar increase (depending on the line 
saturation or position on the ``curve of growth'') in all 
photoexcited lines.  This is in contrast to variations in 
$N^{\mathrm{rad}}_{\mathrm{ion}}$, which primarily 
lead to a ``tilting'' of the photoexcitation contribution, since, one-by-one, 
each line, in descending order of oscillator strength, is eventually 
saturated.  The velocity width we 
invoke here may be due to a turbulent velocity in the plasma or simply due to 
different bulk velocities of multiple absorption systems.

\section{Fits using an Irradiated Cone Model}\label{sec:finalfit}

We have constructed a self-consistent model of an irradiated
cone of plasma for application to the \ngc\ soft-X-ray spectrum.  This
model applies to all significant H-like and He-like ions, employing atomic 
data obtained from the Flexible Atomic Code ({\small FAC}) \cite{gu}.  
(A detailed study of the more complicated Fe~L-shell transitions is left for 
the future.)  To simplify fitting, we have incorporated our model into 
\xspec\ \cite{arnaud} as the local model 
{\it photo}\footnote{http://xmm.astro.columbia.edu/research.html} 
\cite{model}.  

In our model, we allow for individual line absorption, but assume that 
photoelectric absorption is negligible.  With this assumption, we can 
simplify the calculations by irradiating each ionic column density separately 
with an initially unabsorbed power law (discrete line absorption 
due to all other ions can safely be neglected, since 
it removes flux from only a small portion of the total spectrum).
Alternatively,
using the method employed by our model, we can irradiate all ionic column 
densities together, assuming all ions have the same radial distribution.  We 
have checked that both methods produce the same reprocessed spectrum as long 
as all photoelectric edges have low optical depths.  (The radial ionic column 
densities we infer from the spectrum of \ngc\ are indeed consistent with the 
assumption of negligible photoelectric absorption, as we argue below.)  
Furthermore, we assume that the plasma is optically thin to reemitted photons 
(see \S\ref{sec:opticallythin}).  

Our model consists of three free parameters for each ion, the radial velocity 
width $\sigma_v^{\mathrm{rad}}$, the covering factor times nuclear luminosity 
$fL_X$, and the ionic column density $N_{\mathrm{ion}}^{\mathrm{rad}}$.  The 
covering factor, expressed in terms of the solid angle subtended by the 
plasma, is $f=\Omega/4\pi$, and the total power-law luminosity (where, 
throughout, we assume reasonable values for the index of $\Gamma=-1.7$ and 
energy range of 13.6~eV--100~keV) is $L_X$.  In order to convert observed 
flux to 
luminosity, we take the distance to \ngc\ to be 14.4~Mpc 
\cite{bland-hawthorn}.  

We assume the following values for the radial velocity width of 
$\sigma_v^{\mathrm{rad}}=200$~km~s$^{-1}$ and the overall normalization of 
$fL_X=10^{43}$~ergs~s$^{-1}$ are the same for all the ions (discussed further
in \ref{sec:finalfitsub}).  The individual ionic column densities, 
$N_{\mathrm{ion}}^{\mathrm{rad}}$, are left free for each ion.  

The \rrc\ temperatures are directly observed (Table~\ref{tab:rrc}).
Also, the velocity widths of all emission lines transverse to the 
cone are directly observed (Table~\ref{tab:lines}).  For simplicity, we assume 
a velocity width of $\sigma_v^{\mathrm{obs}}\simeq400$~km~s$^{-1}$, which works
well for all of the observed lines.

A more complete discussion of the atomic calculations and astrophysical 
assumptions underlying our model is presented in Kinkhabwala \etal\ (2002).

\subsection{Fit to H-like \ion{C}{6}}

\begin{figure*}[]
\centerline{\psfig{figure=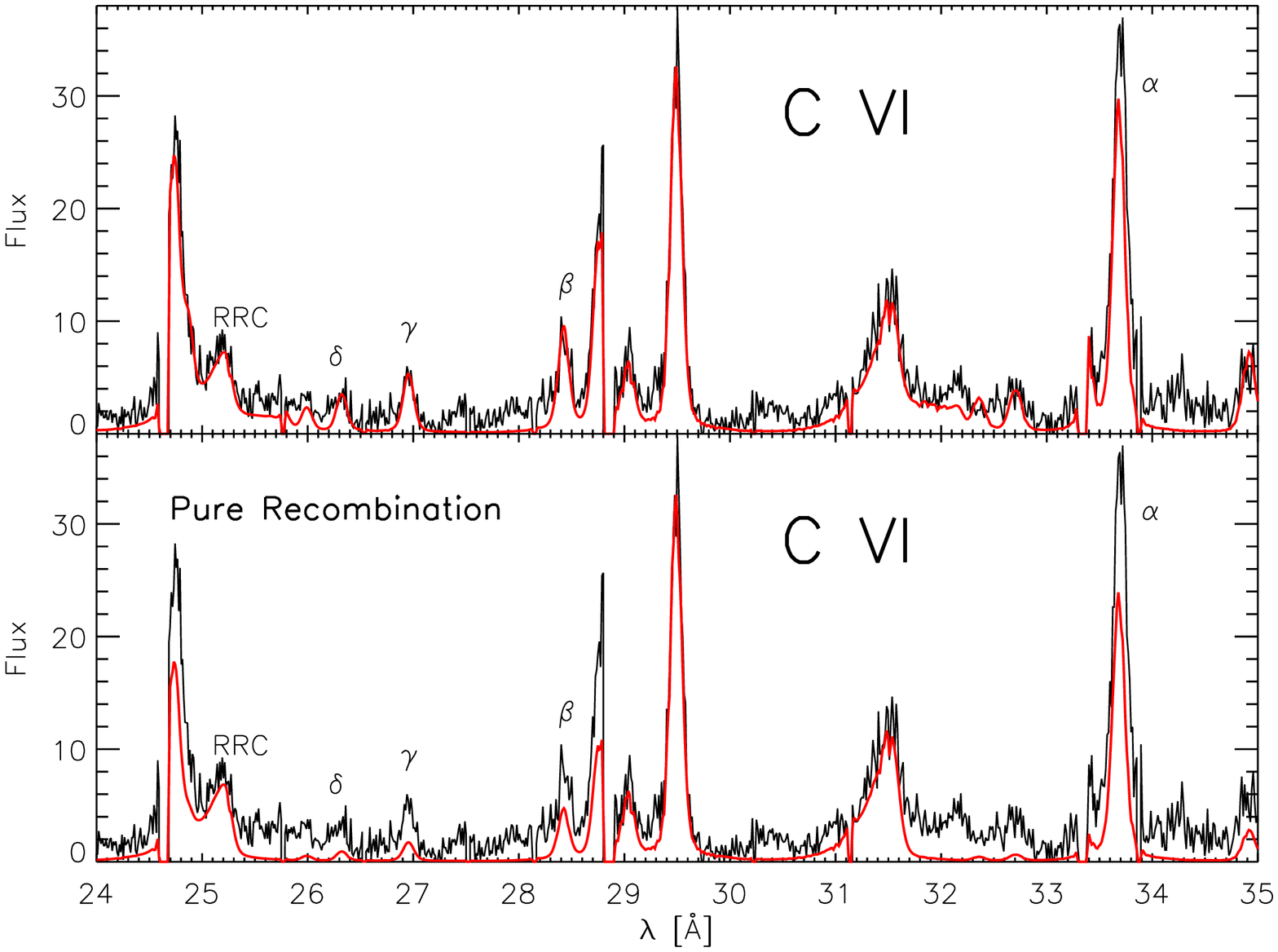,angle=90,width=3.9in}}
\caption{Final fit parameters for C~VI (see Table~\ref{tab:finalfit} and 
Fig.~\ref{fig:finalfit}) including recombination/radiative cascade following 
photoionization and radiative decay following photoexcitation (top).  
Recombination alone (bottom) is unable to explain the excess emission in all 
resonance lines np$\rightarrow$1s.  (Data are plotted in rest frame of 
NGC~1068.  The model spectrum has been velocity shifted by 
$-400$~km~s$^{-1}$.)}
\label{fig:carbon}
\end{figure*}

In Fig.~\ref{fig:carbon}, we show our fit to the line series corresponding to
H-like \ion{C}{6}.  In the bottom panel, we show the best fit possible assuming
only recombination.  Note the gross underprediction of all features relative
to the \rrc.  In the top panel, we self-consistently add in the contribution
from photoexcitation assuming the model described above and a column density
of \ion{C}{6} of $N_{\mathrm{ion}}^{\mathrm{rad}}=9\times10^{17}$~cm$^{-2}$.  
The addition of photoexcitation
completely explains the excess line strengths.

\subsection{Fit to He-like \ion{O}{7}}\label{sec:oxygen}

Similarly, in Fig.~\ref{fig:oxygen}, we show our fit to the line series 
corresponding to He-like \ion{O}{7}.  In the bottom panel, we show the 
best fit possible assuming only recombination.  Note the marked 
underprediction of the resonance line and all higher order features relative 
to the \rrc.  We note that the ratio $R=f/i$ given in Table~\ref{tab:rif} is 
greater than expected for pure recombination.  This may arise from an
additional contribution to the forbidden line due to inner-shell 
photoionization in \ion{O}{6} (e.g., Kinkhabwala et~al. 2002).  In the top
panel, we self-consistently 
add in the contribution from photoexcitation assuming the model described 
above and a column density of \ion{O}{7} of 
$N_{\mathrm{ion}}^{\mathrm{rad}}=1.1\times10^{18}$~cm$^{-2}$.  Again, all the 
higher-order-series lines are completely explained by the addition of 
photoexcitation.

\subsection{Final Fit to All H-like and He-like Ion Series}\label{sec:finalfitsub}

\begin{figure}[htp]
\centerline{\psfig{figure=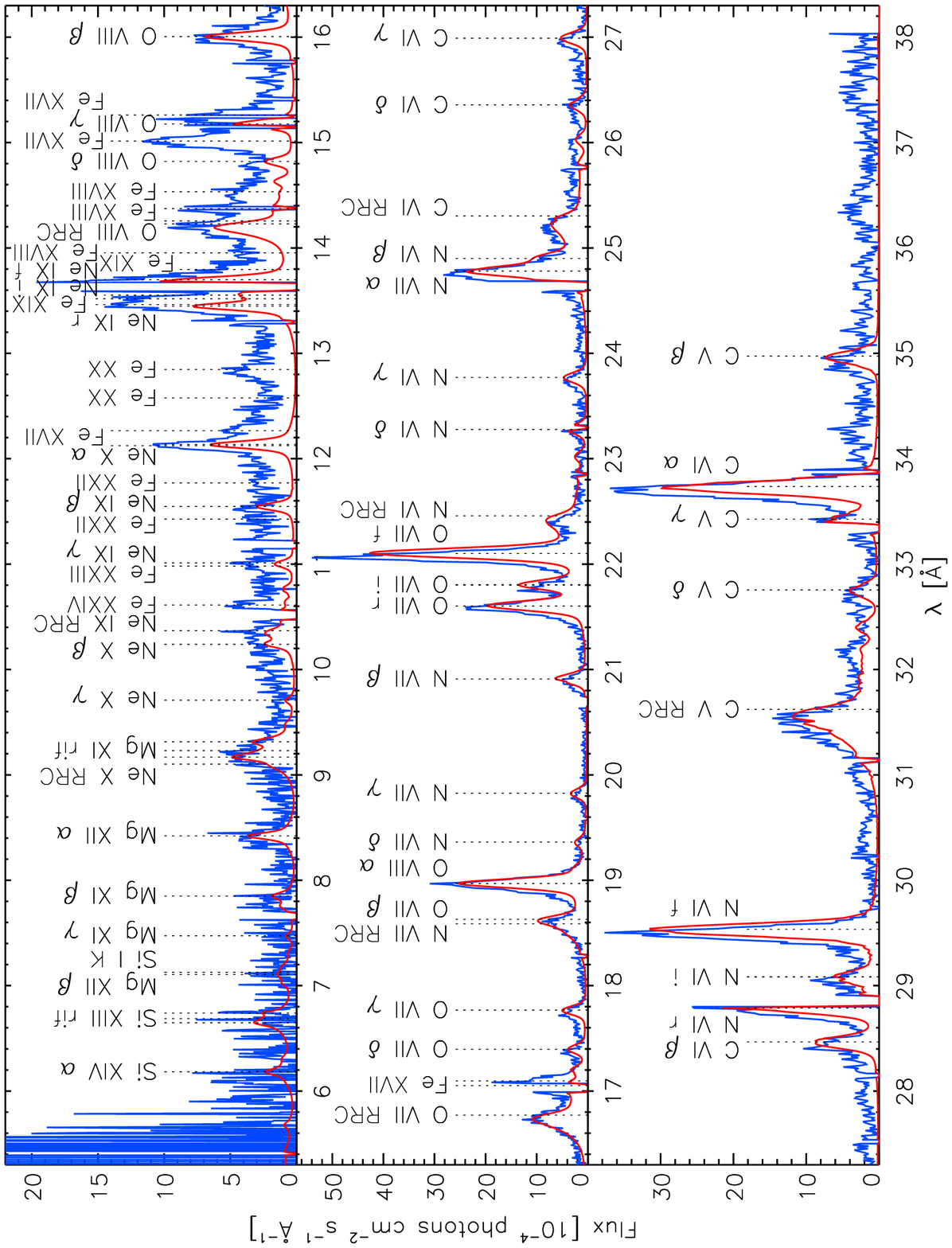,angle=180,width=6.5in}}
\caption{Data from \rgs~1 (blue) plus the best fit model (red)
to NGC~1068, which assumes a normalization of $fL_X=10^{43}$~ergs~s$^{-1}$ and 
radial velocity distribution of $\sigma_v^{\mathrm{rad}}=200$~km~s$^{-1}$, 
with individual ionic column densities given in Table~\ref{tab:finalfit}.  
We assume a total neutral column density to the source of 
$N_{\mathrm H}=3.5\times10^{20}$~cm$^{-2}$ and an 
observed velocity broadening of $\sigma_v^{\mathrm{obs}}=400$~km~s$^{-1}$.  
The data have been been shifted to the rest frame of NGC~1068 ($z=0.00379$).  
The model wavelengths are not shifted.  Line blueshifts, especially 
noticeable at long wavelengths, are clearly present.  Fe~L shell emission 
(spanning 9~\AA\ to 18~\AA) has not yet been included in the model.}
\label{fig:finalfit}
\end{figure}

Our fit to all H-like and He-like ions in the spectrum as a 
whole (neglecting Fe~L shell emission) is presented in 
Fig.~\ref{fig:finalfit}.  By ``fit,'' we simply mean a possible set of 
parameters that can reproduce the spectrum in detail.  Though {\it photo} is a 
local model in \xspec, the interdependent, multicomponent nature of fitting
with this model 
does not all for well-defined error bars for our fitted parameters.  The 
best-fit global values for $\sigma_v^{\mathrm{rad}}$ and $fL_X$ of 
200~km~s$^{-1}$ and $10^{43}$~ergs~s$^{-1}$, respectively, were found by trial
and error.  Attempts were made to fit the spectrum using 
$\sigma_v^{\mathrm{rad}}=100$~km~s$^{-1}$ and 
$\sigma_v^{\mathrm{rad}}=400$~km~s$^{-1}$ with global normalization $fL_X$ and 
individual column densities left free.  However, the details of the 
higher-order transitions for any given ion were less satisfactorily fit.  An 
estimate for the model parameter uncertainty, therefore, gives roughly a 
factor of two for all inferred model parameters: $fL_X$, 
$\sigma_v^{\mathrm{rad}}$, and $N_{\mathrm{ion}}^{\mathrm{rad}}$.  Ratios of 
individual column densities, however, are much better determined (ranging 
from a few percent to a few tens of percent).  Of course, parameters such as 
$\sigma_v^{\mathrm{obs}}$ and the \rrc\ temperatures are directly observed
with error bars quoted in Tables~\ref{tab:lines} and \ref{tab:rrc}.

Assuming $\sigma_v^{\mathrm{rad}}=200$~km~s$^{-1}$ and 
$fL_X=10^{43}$~ergs~s$^{-1}$, as well as the observed broadening of
$\sigma_v^{\mathrm{obs}}=400$~km~s$^{-1}$, we provide
all other best-fit parameters in Table~\ref{tab:finalfit}.  For each 
H-like and He-like ion series, all features (from the 
$\mathrm{n}=2\rightarrow1$ transitions up to the \rrc) are reproduced very 
well (Fig.~\ref{fig:finalfit}) using these parameters.   
The column densities we infer are indeed consistent with our 
assumption of negligible photoelectric absorption and, additionally, are 
consistent (within a factor of a few) with the column densities derived from
several recent high-resolution soft-X-ray spectra of Seyfert~1 galaxies 
\cite{kaastra,kaspi,branduardi-raymont,sakoiras}.  
The inferred radial velocity distribution of 
$\sigma_v^{\mathrm{rad}}=200$~km~s$^{-1}$ is also consistent with the range 
($\sigma_v^{\mathrm{rad}}=100$--600~km~s$^{-1}$) observed in soft-X-ray 
spectra of these same Seyfert~1 galaxies.
Further inspection of Table~\ref{tab:finalfit} reveals that the H-like 
and He-like ions of each element have approximately equal column densities;
this important observation is discussed further in \S\ref{sec:ksidist}. 
Assuming Solar abundances 
(Grevesse, Noels, \& Sauval 1996; Allende Prieto, Lambert, \& Asplund 2001) 
and a fractional ionic abundance of \ion{O}{8} of $f_{\mathrm{i}}=0.5$, we 
obtain a column density in H of $N_{\mathrm{H}}\gsim4\times10^{21}$~cm$^{-2}$ 
(see \S\ref{sec:column}).  

We also give measurements of the emission measure of the recombining plasma
(${\mathrm{EM}}=\int n_{\mathrm{e}}^2dV$) for each ion in 
Table~\ref{tab:finalfit}.  For all {\small EM} values, we assume that the 
recombining ion fractional abundance is near 
maximal at $f_{i+1}=0.5$.  For C, N, and O, these emission measures are based 
on the measured \rrc\ temperatures and fluxes.  For Ne, Mg, and Si, the 
{\small EM} is based on the arbitrarily chosen temperature of 
$kT_{\mathrm{e}}=4$~eV.  We note that the temperatures for these ions are not 
likely to be lower than $kT_{\mathrm{e}}=4$~eV, but could easily be an order 
of magnitude higher (see \S\ref{sec:ksi}), implying the {\small EM}s we list 
are really only lower limits for these ions.

\section{Discussion}\label{sec:dis}

We have shown that the soft-X-ray spectrum of the prototypical Seyfert~2 
galaxy \ngc\ is dominated by emission from warm plasma irradiated by the 
nuclear continuum.  In addition, we have shown that the relative amounts of 
photoionization versus photoexcitation for all ionic transition 
series provide detailed, self-consistent information about the geometry and 
dynamics of the 
emission regions, yielding radial column densities, radial velocity widths, 
and the product of covering factor and nuclear luminosity.  The values we 
obtain for these geometrical properties (in particular, the radial ionic 
column densities) are remarkably similar to absorption measurements
in Seyfert~1 galaxies.  We have also shown that there is no measurable 
additional component due to hot, collisional plasma (e.g., from shocks, a 
circumnuclear 
starburst, or a hot confining medium).  
We can put a robust upper limit to an additional hot, 
collisional plasma component in \ngc\ at an order-of-magnitude less than the
line luminosities measured in the soft-X-ray spectrum, or 
${\mathrm{EM}}=\int n_{\mathrm e}^2 dV\leq 3\times10^{62}$~cm$^{-3}$ 
for temperatures in the range $kT=400$--900~eV.  In the following, we
explore further relevant astrophysical considerations.

\subsection{Radial Filling Factor of the Warm Absorbing Plasma}\label{sec:warmabsorber}

\begin{figure*}[]
\centerline{\psfig{figure=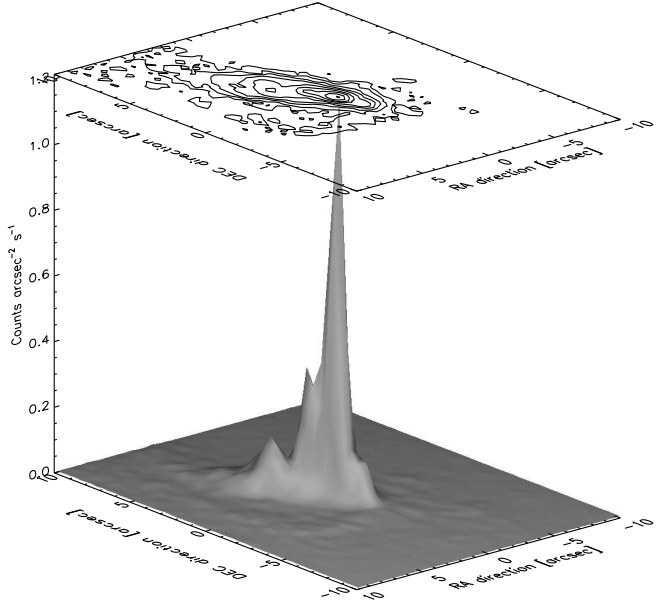,angle=90,width=4.6in}}
\caption{{\it Chandra} ACIS-S contour image of NGC~1068 using a frame time of 
0.4 s and total exposure of 11465.4 s (Young \etal\ 2001).  Only X-rays with 
CCD-determined energies less than 2.5 keV were taken.  The image is centered 
on the peak of the X-ray emission at 
$\alpha$(J2000)=02$^{\mathrm{h}}$42$^{\mathrm{m}}$40$^{\mathrm{s}}$.71, 
$\delta$(J2000)=$-$00$^{\circ}$00\arcmin47\arcsec.9 ($1\arcsec=72$~pc).  The 
position of the peak of the 5~GHz radio emission (assumed nuclear position) at 
$\alpha$(J2000)=02$^{\mathrm{h}}$42$^{\mathrm{m}}$40$^{\mathrm{s}}$.715, 
$\delta$(J2000)=$-$00$^{\circ}$00\arcmin47\arcsec.64 \cite{muxlow} is within
0\arcsec.3 of the X-ray peak, which is well within the {\it Chandra} 
positional uncertainty.  We use factor-of-two contours with the lowest at 
$4.4\times10^{-3}$~counts~arsec$^{-2}$~s$^{-1}$.  The cone is clearly 
extended over hundreds of parsec.}
\label{fig:contour_acis}
\end{figure*}

The \chandra\ image of \ngc\ shown in Fig.~\ref{fig:contour_acis} reveals 
that a large fraction of the soft-X-ray emission from the ionization cone 
comes from a region close to the nucleus, as 
inferred from radio observations.  The minimum radius of the ionization cone
is clearly $\lsim100$~pc, but could be much smaller.  It is possible that
a large fraction
of the cone is obscured from our view; however, the observed 
photoexcitation of low-n, 
high-oscillator-strength transitions implies that we must be directly viewing 
the bulk of the ionization cone, since these high-oscillator-strength 
transitions saturate very quickly (see Fig.~\ref{fig:opticaldepth}).  This sets
a lower limit on the radius of the bulk of the ionization cone which must
be greater than the radius of the obscured circumnuclear region.  

Using upper limits to the covering factor ($f\lsim0.1$) and minimum
radius ($r_{\mathrm{min}}\lsim100$~pc), we can derive an upper
limit to the radial filling factor of the plasma $g$, which, by definition,
is $\leq1$.
We can estimate the column density for a given ion as follows:
\begin{eqnarray}
\lefteqn{N^{\mathrm{rad}}_{\mathrm{ion}}=\int_{r_{\mathrm{min}}}^{\infty}\frac{A_Z f_i}{1.2} g(r)\frac{L_X}{\xi r^2}dr\simeq1.6\times10^{24}A_Z\times}\nonumber\\
& \frac{g}{fr_{\mathrm{min,pc}}}\big[\frac{f_i}{0.5}\big]\big[\frac{(fL_X)}{10^{43}~\mathrm{ergs~s}^{-1}}\big]\big[\frac{\xi}{1~\mathrm{ergs~cm~s}^{-1}}\big]^{-1}~\mathrm{cm}^{-2},
\end{eqnarray}
where $A_Z$ is the elemental abundance, $f_i$ is the fractional ionic 
abundance, $r_{\mathrm{min,pc}}$ is the minimum radius of the cone measured 
in parsec, $g(r)$ is the radial filling as a function of radius, and 
$g$ is a single number representing the average radial filling in the cone.  
Assuming the relevant values for \ion{C}{5} of 
$A_Z=3.63\times10^{-4}$ (Grevesse et~al. 1996), 
$\xi=1$~erg~cm~s$^{-1}$, and observed column density of 
$N^{\mathrm{rad}}_{\mathrm{ion}}=8\times10^{17}$~cm$^{-2}$, we obtain 
$g = 1.4\times10^{-3}fr_{\mathrm{min,pc}}$.  Similarly, assuming
the relevant values for \ion{O}{8} of $A_Z=4.89\times10^{-4}$ (Allende Prieto 
et~al. 2001), $\xi=10$~ergs~cm~s$^{-1}$, and observed column density of 
$N^{\mathrm{rad}}_{\mathrm{ion}}=1\times10^{18}$~cm$^{-2}$, we obtain 
$g = 1.3\times10^{-2}fr_{\mathrm{min,pc}}$.  Using robust
upper limits for the covering factor and the minimum radius from the Chandra 
image (Fig.~\ref{fig:contour_acis}) of $f\lsim0.1$ and 
$r_{\mathrm{min,pc}}\lsim100$~pc, we obtain upper limits of 
$g\lsim1.4\times10^{-2}$ (plasma at $\xi=1$~erg~cm~s$^{-1}$) and 
$g\lsim1.3\times10^{-1}$ (plasma at $\xi=10$~ergs~cm~s$^{-1}$).  The actual
filling factor, therefore, is significantly less than unity, 
suggesting that the X-ray plasma fills only a small portion of the ionization 
cone.  

The number densities for plasma with ionization parameter 
$\xi=1$~erg~cm~s$^{-1}$ at 500~pc (maximum extent of cone from 
Fig.~\ref{fig:contour_acis}) down to 1~pc (arbitrarily-assumed minimum 
radius), taking $L_X=10^{44}$~ergs~s$^{-1}$, are
$n_{\mathrm{e}}=4\times10^1$--$1\times10^7$~cm$^{-3}$.  These densities are a 
factor of ten less for plasma with $\xi=10$~ergs~cm~s$^{-1}$.

\subsection{Lack of Electron-scattered Continuum and $N_{\mathrm{H}}$}\label{sec:column}

We observe no significant electron-scattered continuum.  This allows us to 
obtain an upper limit to the electron column density in the context of our 
model as follows.  From our measured value for the covering factor times 
nuclear luminosity of $fL_X=10^{43}$~ergs~s$^{-1}$ and the assumed distance to 
\ngc\ of 14.4~Mpc, we can write the column density in electrons as a function 
of the normalization $A$ of the reflected continuum 
$F_{\mathrm{refl}}=AE^{-1.7}$ ($F_{\mathrm{refl}}$ has units of 
photons~cm$^{-2}$~s$^{-1}$~keV$^{-1}$ with $E$ in keV) as 
$N_{\mathrm{e}}=1.0\times 10^{26} A$~cm$^{-2}$.  The observed upper limit to 
the electron-reflected continuum of $A\lsim8\times10^{-4}$ (using the 
line-free region between 20.0--20.6~\AA) gives 
$N_{\mathrm{e}}\lsim8\times10^{22}$~cm$^{-2}$.  We have already found
a reasonable lower limit to $N_{\mathrm{H}}$ of $4\times10^{21}$~cm$^{-2}$
in \S\ref{sec:finalfitsub}.  Therefore, assuming H and He are fully stripped, 
we constrain the neutral column density to be 
$4\times10^{21}$~cm$^{-2}\lsim N_{\mathrm{H}}\lsim7\times10^{22}$~cm$^{-2}$.

\subsection{Optical Depth in Transverse Direction}\label{sec:opticallythin}

The optical depth in a given line for an approximately spherical 
cloud determines the average number of scatters a line photon will experience
before it is either destroyed or escapes.  
For significant column density transverse to the cone (i.e.,
along the Seyfert~2 line of sight in Fig.~\ref{fig:cone}), the
higher-order resonance transitions (np$\rightarrow$1s with $\mathrm{n}>2$) and 
photons created by recombination directly to ground will be reprocessed and 
degraded to $\mathrm{n}=2\rightarrow1$ transitions \cite{bowen}.

Since we do not know the geometry of the scattering regions (e.g., filled 
cone or isolated clouds), we employ order-of-magnitude estimates.   If 
$N_{\mathrm{ion}}^{\mathrm{cloud}}$ is the 
ionic column density and $\sigma_{\mathrm{ave}}$ is the line-profile-averaged 
cross section for a particular transition (taking into account velocity 
broadening $\sigma_v$) in a roughly spherical cloud, then the optical 
thickness is $\tau=N_{\mathrm{ion}}^{\mathrm{cloud}}\sigma_{\mathrm{ave}}$ 
and the average number of scatters is $N\approx\tau$.
Applying these considerations to H-like and He-like series, we find 
that conversions of Ly$\beta$ to Ly$\alpha$ and He$\beta$ to 
He$\alpha$ provide the most sensitive diagnostics.  The branching ratios
are roughly 0.9 to produce the same photon and therefore 0.1 to cascade 
differently.

\begin{figure*}[]
\centerline{\psfig{figure=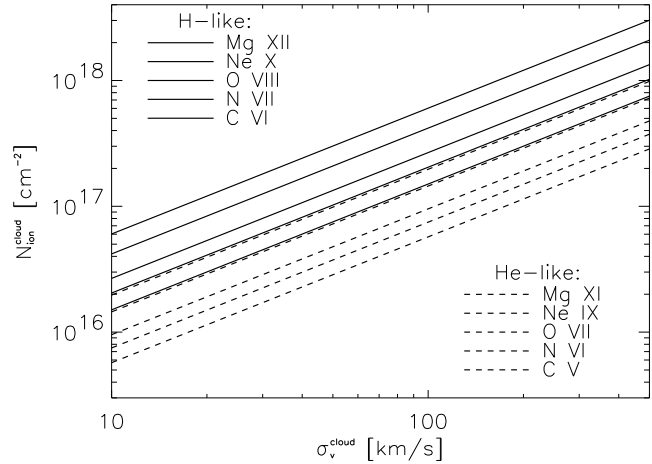,angle=90,width=3.7in}}
\caption{Upper limits to ionic column densities assuming 10\%\ 
conversion percentage of Ly$\beta$ photons to Ly$\alpha$
and, similarly, He$\beta$ to He$\alpha$ due to multiple scatterings in
a roughly isotropic medium.  Depending on the filling factor, the medium 
can be thought of as either an individual cloud or as the entire emission 
region.  (In order to further generalize these results for arbitrary 
line conversion, each line in the plot can be multiplied by 
$\ln{f}/\ln{0.9}$, where $1-f$ is the degradation of the specific
3p$\rightarrow$1s transition.)  The labels in the upper left and lower right 
reflect the order of the lines, from top to bottom, in the plot.    Robust 
upper and lower limits to $\sigma_v$ of 10~km~s$^{-1}$ and 500~km~s$^{-1}$
are from thermal broadening ($kT=3$~eV) and the observed emission line 
broadening, respectively.  For H-like C~VI and He-like C~V, N~VI, and 
O~VII, the upper limits over the entire range are consistent with or 
significantly less than the fitted radial column densities 
(Table~\ref{tab:finalfit}), possibly suggesting a narrow cone geometry.}
\label{fig:escape}
\end{figure*}

Our final fit to the spectrum of \ngc\ shown in Fig.~\ref{fig:finalfit} 
assumes that
all reemission occurs in optically-thin plasma, i.e., all photons created in 
the plasma escape with no further reprocessing.  Therefore, due to the goodness
of the fit, we assume at most a 
10\% reduction (roughly the limit of detectability) in the Ly$\beta$ and 
He$\beta$ lines, we plot
the upper limit to the ionic optical depth versus $\sigma_v$ for several
different ions in Fig.~\ref{fig:escape}.  Robust lower and upper limits to 
$\sigma_v$ of 10~km~s$^{-1}$ and 500~km~s$^{-1}$ are set by thermal broadening 
($kT\sim3$~eV) and the observed line broadening, respectively.
He-like ions provide the best upper limits to ionic column densities.  
Notice that
the upper limits (over the entire range in $\sigma_v$) to the ionic column
densities of \ion{C}{5}, \ion{N}{6}, and \ion{O}{7} are 
consistent-with-to-lower-than the inferred column
densities along the cone given in Table~\ref{tab:finalfit} (where we have
assumed a radial velocity distribution of 
$\sigma_v^{\mathrm{rad}}=200$~km~s$^{-1}$).  This may
suggest that the radial column density is larger by roughly a factor of a few
than the transverse column density, which would be consistent with a narrow 
cone geometry.

\subsection{Electron Temperatures}\label{sec:ksi}

\begin{figure*}[]
\centerline{\psfig{figure=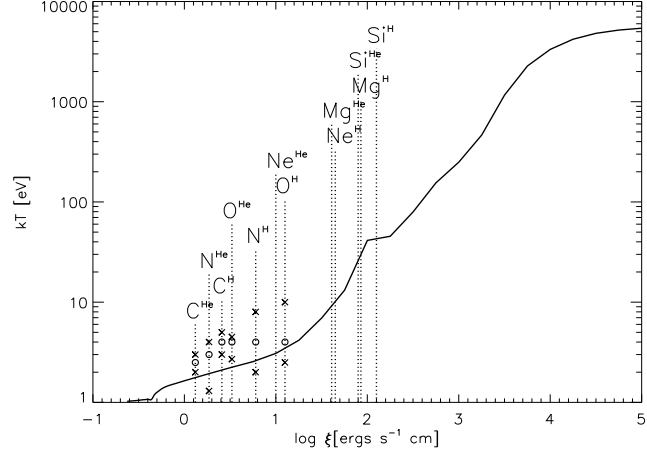,angle=90,width=3.5in}}
\caption{From a photoionization simulation using XSTAR of an optically-thin,
low-density cloud of Solar-abundance plasma surrounding an ionizing point 
source with power-law spectrum ($\Gamma=1.7$, $E_{\mathrm{min}}=13.6$~eV, 
$E_{\mathrm{max}}=100$~keV), we have determined the expected 
relationship between temperature $kT$ [eV] and ionization parameter 
$\xi=L_X/(n_{\mathrm e} r^2)$ (solid curve), where $L_X$  [ergs s$^{-1}$] is 
the total luminosity in the power-law continuum, $n_{\mathrm e}$ 
[cm$^{-3}$] is the electron density, and $r$ [cm] is the distance from the
nucleus to the emission region.  Peaks in the recombination emissivity forming
H-like and He-like ions are indicated (dotted lines).  The FWHM for 
each ion's emissivity distribution is $\Delta(\log{\xi})\approx1$.  The  x's 
and $\circ$'s denote the temperature confidence interval and best fit, 
respectively, taken directly from Table~\ref{tab:rrc}.}
\label{fig:ksi}
\end{figure*}

The electron temperatures we measure directly from the \rrc\ for recombination
onto specific ions can be compared with results from self-consistent 
simulations of photoionized plasma, where photoelectric heating is balanced by 
line cooling in the plasma.  Using the photoionization code \xstar\ 
\cite{kallman}, we have determined the expected temperature, 
$kT$, versus ionization parameter, $\xi=L_X/n_e r^2$, for a spherical, 
optically-thin plasma with central, ionizing, power-law continuum 
($\Gamma=-1.7$, $E_{\mathrm{min}}=13.6$~eV, $E_{\mathrm{max}}=100$~keV).  In 
Fig.~\ref{fig:ksi}, we present the expected range in 
temperature and ionization parameter for all observed H-like and 
He-like ions.  The \xstar-predicted temperatures agree well with 
the data.

\subsection{Thermal Instability}\label{sec:instability}

\begin{figure*}[]
\centerline{\psfig{figure=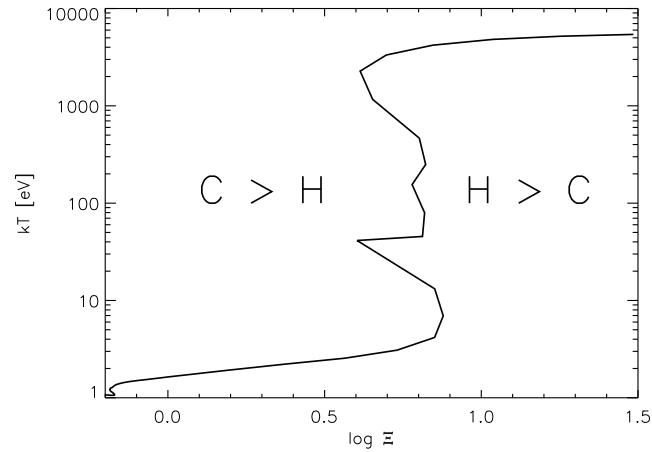,angle=90,width=3.5in}}
\caption{For the same simulation as in Fig.~\ref{fig:ksi}, we have
determined the curve of thermal stability using $\Xi=\xi/4\pi kT$ 
(Krolik, McKee, \& Tarter 1981).  Regions along the thermal 
equilibrium curve that have negative slope are thermally unstable (cooling is 
greater than heating for regions leftward of the curve and vice 
versa for regions rightward).  Several ions observed in the spectrum
of NGC~1068 have maximal abundance (see Fig.~\ref{fig:ksi}) in the thermally 
unstable region.}
\label{fig:Ksi}
\end{figure*}

For a generic photoionized plasma, thermal instabilities may develop.
In Fig.~\ref{fig:Ksi}, we plot the thermal equilibrium curve for the same
optically thin plasma (assuming Solar abundances) along which heating equals 
cooling.  The plasma is thermally unstable along segments of the 
equilibrium curve which have negative slope, since upward deviations from the 
curve in this region heat
the plasma until it reaches the higher stable branch and downward deviations
cool the plasma until it reaches the lower stable branch.  Comparing both 
Figs.~\ref{fig:ksi} and \ref{fig:Ksi}, we see that several ions present in the 
spectrum have maximal ionic abundances in this thermally unstable region.  
This putative thermal instability, therefore, appears problematic
for ions in the observed range of ionization parameter, $\xi$.  

We note that thermal instability can be removed if additional heating/cooling 
mechanisms are at work in the plasma in addition to those associated with 
photoelectric heating and line cooling.  Thermal instability is also 
sensitive to metal abundance (Hess, Kahn, \& Paerels 1997), with low metal 
abundances effectively removing the instability.  The existence of thermal 
instability in astrophysical photoionized plasmas is presently an open 
question.

\subsection{Ionization Parameter Distribution}\label{sec:ksidist}

In Fig.~\ref{fig:fractional}, we present the fractional ionic abundances 
relative to H and assuming Solar abundances for the elements observed in 
the \rgs\ spectrum of \ngc.  No single ionization parameter
is capable of explaining the column densities obtained in 
Table~\ref{tab:finalfit}, which shows similar column densities in H-like 
and He-like ions for each elemental species.  The data instead require a 
broad, flat distribution in ionization parameter of the plasma over the 
range $\log{\xi}\sim0$--3.  This distribution in $\xi=L_X/n_{\mathrm{e}}r^2$ 
could be due to spatial stratification of ionization zones in a 
single-density cone \cite{sakocircinus} or a broad density distribution
at each radius.  

Kraemer and Crenshaw (2000b) find that
the extended optical/\uv\ absorber is powered by photoionization due to an 
inferred
hidden source of continuum, as we similarly argue for the soft-X-ray 
emission.  These findings, coupled with the remarkable 
overlap of the soft-X-ray image of the cone with optical observations of 
\ion{O}{3} emission regions (Young \etal\ 2001), favor an intrinsic 
density 
distribution (over at least three orders of magnitude) at each radius.
For such a model, a perfectly flat distribution in $\log{\xi}$ would give a 
distribution in densities at each radius of 
$f(n_{\mathrm{e}})\propto n_{\mathrm{e}}^{-1}$.  
An intrinsic density distribution in \agn\ ionization cones has been 
proposed in the past to explain the optical and \uv\ spectra of generic 
Seyfert~2 galaxies (e.g., Komossa \& Schulz 1997).  In X-ray emitting 
plasmas, such a density distribution may naturally arise in outflows due to 
thermal instabilities (Krolik \& Kriss 2001), which may counter the usual
perception of such instabilities on the presence of specific ions 
in the unstable region (see \S\ref{sec:instability}).

The upper limits obtained in \S\ref{sec:warmabsorber} for the radial filling
factor do not allow for distinguishing between spatially-stratified 
ionization zones or an intrinsic density distribution.  Robust discrimination
between these two models will require future observations and analysis.

\subsection{Relative Elemental Abundances}

Given the ionization parameter distribution, the inferred ionic column 
densities can provide estimates of relative elemental abundances.
However, even absent knowledge of the exact distribution, some tentative 
conclusions can be drawn due to the overlap in ionization parameter of 
specific charge states of different ions.

\begin{figure*}[]
\centerline{\psfig{figure=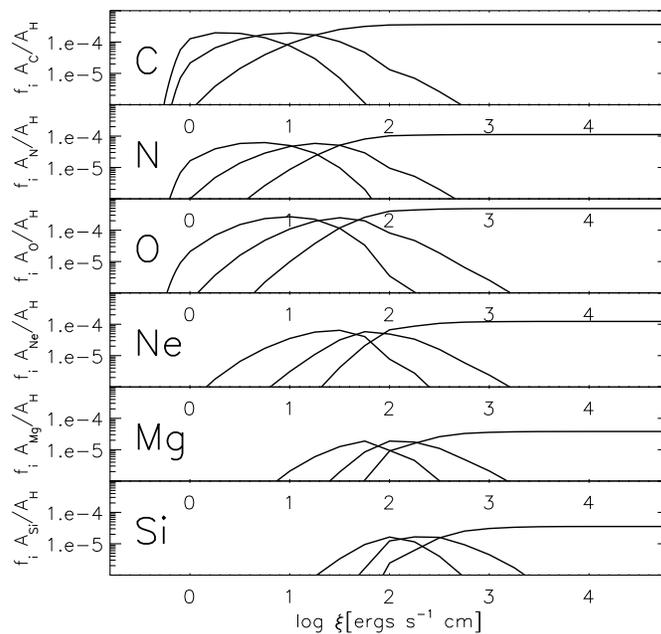,angle=90,width=4.5in}}
\caption{Fractional ionic abundances relative to H (assuming Solar abundances) 
for He-like (leftmost curves), H-like (middle), and bare (rightmost) ions 
using the same XSTAR simulation as in Figs.~\ref{fig:ksi} and ~\ref{fig:Ksi}.  
The observed similar column densities in the H-like and He-like charge 
states of each ion (Table~\ref{tab:finalfit}) require a broad distribution in 
$\xi$.  Assuming relative Solar abundances all ions, the observation of 
column densities in C~V and C~VI of 8--9$\times10^{17}$~cm$^{-2}$ and in 
Si~XIII and Si~XIV of $2\times10^{17}$~cm$^{-2}$ requires this distribution to 
be rather flat as well, spanning the range $\log\xi=0$--3.}
\label{fig:fractional}
\end{figure*}

The necessity of a broad, rather flat range of ionization parameter allows
for comparison of overlapping charge states of different elements, 
such as H-like C and He-like O (see Fig.~\ref{fig:fractional}).  Overall, we 
find that all of the
well-measured ionic column densities yield relative elemental abundances which 
are consistent with Solar (Grevesse et~al. 1996;
 Allende Prieto et~al. 2001), excluding N, which is overabundant by 
approximately a factor of three.  The column densities of Mg and 
Si also seem high in the context of a flat ionization parameter distribution.  
But, due to low counts and blending, their column densities are not well 
determined, making difficult any robust conclusions concerning their relative
abundance.

Furthermore, we note that the abnormally high Fe/O abundance invoked for 
previous \asca\ spectra of \ngc\ by Netzer \& Turner (1997) is not observed.
In fact, the \ion{O}{7} and \ion{O}{8} emission lines are the strongest in the 
spectrum, especially when 
compared to Fe L-shell transitions, which are considerably weaker (but not 
necessarily inconsistent with Solar-abundance photoionized plasma).  Fe 
L-shell line emission in generic photoionized plasma, as well as in \ngc, is 
currently under investigation.

\acknowledgments{
We would like to thank Julian Krolik and the anonymous referee for helpful 
comments.  {\small AK} acknowledges useful discussions with all members of 
the Columbia \xmm\ \rgs\ group.  This work is based on observations obtained 
with \xmm, an \esa\ science mission with instruments and contributions 
directly funded by \esa\ Member States and the \usa\ (\nasa).  The Columbia 
University team is supported by \nasa.  {\small AK} acknowledges additional 
support from an \nsf\ Graduate Research Fellowship and \nasa\ \gsrp\ 
fellowship.  {\small MS} and {\small MFG} were partially supported by 
{\small NASA} through {\it   Chandra} Postdoctoral Fellowship Award Numbers 
{\small PF}01-20016 and {\small PF}01-10014, respectively, issued by the 
{\it Chandra} X-ray Observatory Center, which is operated by the Smithsonian 
Astrophysical Observatory for and behalf of {\small NASA} under contract 
{\small NAS}8-39073.  \sron\ is supported by the Netherlands Foundation for 
Scientific Research (\nwo).  Work at LLNL was performed under the auspices of 
the U. S. Department of Energy, Contract No. W-7405-Eng-48.
}


\bibliographystyle{apj}

\newpage

\newpage

\newpage

\newpage

\newpage

\newpage

\newpage

\newpage

\newpage

\newpage

\newpage

\newpage

\newpage

\newpage

\newpage

\clearpage
\newpage
\begin{deluxetable}{lcccc}
\tablecolumns{5}
\tablewidth{6.8in}
\tablehead{\colhead{Line {\small ID}} & \colhead{$\lambda_{\mathrm{expected}}$[\AA]}   & \colhead{Shift [km~s$^{-1}$]\tablenotemark{a}} & \colhead{$\sigma_v^{\mathrm{obs}}$
[km~s$^{-1}$]\tablenotemark{a}}  & \colhead{Flux [$10^{-4}$~photons~cm$^{-2}$~s$^{-1}$]\tablenotemark{a}}}
\startdata
\ion{Fe}{20} 			&12.846		&$-500\pm190$	& --- 		&$0.60\pm0.15$ \\
\hline 
\ion{Fe}{17}			&15.014		&$-60\pm160$	&$570\pm100$\tablenotemark{b} 	&$1.67\pm0.17$ \\
\ion{Fe}{17}			&15.261		&$-210\pm160$	&$570\pm100$\tablenotemark{b} 	&$0.65\pm0.13$ \\
\ion{Fe}{17} (2)	&17.051,17.096	&$-180\pm140$	& --- 		&$2.78\pm0.28$ \\
\hline
\ion{Si}{14}   Ly$\alpha$	&6.18223	&---		& --- & $\lsim0.8\pm0.2$ \\
\hline
\ion{Si}{13}  rif		&6.69		&---		& --- & $\lsim0.8\pm0.2$ \\
\hline
\ion{Si}{1}  K			&7.1542		&---		& --- & $\lsim0.8\pm0.2$ \\
\hline
\ion{Mg}{12}   Ly$\alpha$	&8.42100	&$70\pm250$	& --- & $0.32\pm0.06$ \\
\hline
\ion{Mg}{11}    rif 		&9.23		&---	    	& --- & $0.26\pm0.10$ \\
\hline
\ion{Ne}{10}	Ly$\alpha$ 	&12.134  	&$-100\pm200$	&$470\pm120$ 	&$1.47\pm0.15$ \\
\hline
\ion{Ne}{9}    	He$\gamma$	&11.000		&$110\pm220$	& ---  		&$0.33\pm0.06$ \\
\ion{Ne}{9}    	He$\beta$	&11.547		&---		& ---  		&$\lsim0.42\pm0.07$\\
\ion{Ne}{9}	r		&13.447		&$-90\pm180$	& --- 		&$1.59\pm0.17$\\
\ion{Ne}{9}	i		&13.552		&---		& --- 		&$\lsim0.83\pm0.17$\\
\ion{Ne}{9}	f		&13.698		&$-30\pm180$	& --- 		&$2.03\pm0.20$ \\
\hline
\ion{O}{8}   	Ly$\delta$	&14.821		&---	        & ---		&$\lsim0.37\pm0.06$\\
\ion{O}{8}   	Ly$\beta$	&16.006		&$-90\pm150$	& --- 		&$1.45\pm0.15$ \\
\ion{O}{8}	Ly$\alpha$  	&18.969	  	&$-200\pm130$	&$700\pm80$ 	&$5.89\pm0.59$ \\
\hline
\ion{O}{7}    	He$\delta$	&17.396		&$-520\pm140$	& --- 		&$0.63\pm0.07$ \\
\ion{O}{7}    	He$\gamma$	&17.768		&$-380\pm140$	&$640\pm80$ 	&$0.79\pm0.08$ \\
\ion{O}{7}    	He$\beta$	&18.627		&$-430\pm130$	& ---		&$1.32\pm0.13$ \\
\ion{O}{7}	r	 	&21.602	  	&$-260\pm110$	&$450\pm70$\tablenotemark{b}	&$4.50\pm0.45$ \\
\ion{O}{7}      i		&21.803		&$-380\pm110$	&$450\pm70$\tablenotemark{b}	&$1.75\pm0.19$ \\
\ion{O}{7}	f	 	&22.101	 	&$-430\pm110$	&$450\pm70$\tablenotemark{b}	&$9.59\pm0.96$ \\
\hline
\ion{N}{7}    	Ly$\delta$	&19.361		&$-380\pm120$	& --- 		&$0.12\pm0.06$ \\
\ion{N}{7}    	Ly$\gamma$	&19.826		&$-350\pm120$	&$390\pm80$	&$0.44\pm0.07$ \\
\ion{N}{7}    	Ly$\beta$	&20.910		&$-370\pm120$	&$520\pm70$	&$0.81\pm0.12$ \\
\ion{N}{7}    	Ly$\alpha$ 	&24.781  	&$-270\pm100$	&$650\pm60$	&$6.09\pm0.74$ \\
\hline
\ion{N}{6}    	He$\delta$	&23.277 	&$-210\pm100$	&$470\pm60$	&$0.32\pm0.14$ \\
\ion{N}{6}    	He$\gamma$	&23.771		&$-490\pm100$	&$360\pm60$	&$0.77\pm0.12$ \\
\ion{N}{6} 	r 		&28.787	 	&$-350\pm80$	&$410\pm50$\tablenotemark{b}	&$3.84\pm0.38$ \\
\ion{N}{6} 	i		&29.083		&$-310\pm80$	&$410\pm50$\tablenotemark{b}	&$1.22\pm0.15$ \\
\ion{N}{6} 	f	 	&29.534  	&$-430\pm80$	&$410\pm50$\tablenotemark{b}	&$8.46\pm0.85$ \\
\hline
\ion{C}{6} 	Ly$\delta$ 	&26.357 	&$-440\pm90$	&$400\pm60$	&$0.57\pm0.10$ \\
\ion{C}{6} 	Ly$\gamma$ 	&26.990 	&$-440\pm90$	&$320\pm60$	&$0.95\pm0.11$ \\
\ion{C}{6} 	Ly$\beta$ 	&28.466		&$-380\pm80$	&$450\pm50$	&$2.03\pm0.20$ \\
\ion{C}{6} 	Ly$\alpha$ 	&33.736  	&$-360\pm70$	&$510\pm40$  	&$12.29\pm1.23$	\\
\hline
\ion{C}{5}      He$\delta$	&32.754		&$-560\pm70$	&$510\pm50$ 	&$1.33\pm0.20$\\
\ion{C}{5}      He$\gamma$	&33.426		&  ---		&$550\pm50$  	&$1.50\pm0.31$\\
\ion{C}{5}      He$\beta$	&34.973		&$-550\pm70$	&$360\pm40$ 	&$1.28\pm0.30$\\
\enddata
\tablenotetext{a}{Error bars (due to dominant systematic instrumental 
uncertainty) are 1-$\sigma$.}
\tablenotetext{b}{Fe L-shell lines and He-like triplet linewidths for N and O were tied together for fitting for each line complex.}
\label{tab:lines}
\end{deluxetable}

\newpage
\begin{deluxetable}{lcc}
\tablecolumns{3}
\tablewidth{4.1in}
\tablehead{\colhead{Ion} & \colhead{\rrc\ $kT_{\mathrm{e}}$ [eV]\tablenotemark{a}} & \colhead{Flux [$10^{-4}$~photons~cm$^{-2}$~s$^{-1}$]\tablenotemark{b}}}
\startdata
\ion{C}{5}	& $2.5_{-0.5}^{+0.5}$ & $4.30\pm0.43$\\
\ion{C}{6}	& $4.0_{-1.0}^{+1.0}$ & $2.83\pm0.28$\\
\ion{N}{6}	& $3.0_{-2.0}^{+1.0}$ & $2.06\pm0.21$\\
\ion{N}{7}	& $4.0_{-2.0}^{+4.0}$ & $1.14\pm0.11$\\
\ion{O}{7}	& $4.0_{-1.3}^{+0.5}$ & $2.43\pm0.24$\\
\ion{O}{8}	& $4.0_{-1.5}^{+6.0}$ & $1.25\pm0.13$\\
\enddata
\tablenotetext{a}{Error bars denote approximate 90\% confidence interval.}
\tablenotetext{b}{Error bars (due to dominant effective-area uncertainty) 
are 1-$\sigma$.}
\label{tab:rrc}
\end{deluxetable}

\newpage
\begin{deluxetable}{lcccc}
\tablecolumns{5}
\tablewidth{4.in}
\tablehead{\colhead{Ion} & \colhead{Line Ratio} & \colhead{\ce \tablenotemark{a}}& \colhead{\rec \tablenotemark{b}} & \colhead{Observed\tablenotemark{c}}}
\startdata
\ion{C}{6} & Ly$\beta/$Ly$\alpha$ & 0.090 & 0.144 & $0.165\pm0.023$\\
           & Ly$\gamma/$Ly$\alpha$& 0.024 & 0.052 & $0.077\pm0.012$\\
	   & Ly$\delta/$Ly$\alpha$& 0.010 & 0.026 & $0.046\pm0.009$\\
\ion{N}{7} & Ly$\beta/$Ly$\alpha$ & 0.096 & 0.141 & $0.133\pm0.025$\\
           & Ly$\gamma/$Ly$\alpha$& 0.023 & 0.049 & $0.071\pm0.014$\\
	   & Ly$\delta/$Ly$\alpha$& 0.009 & 0.024 & $0.020\pm0.010$\\
\ion{O}{8} & Ly$\beta/$Ly$\alpha$ & 0.100 & 0.138 & $\lsim0.247\pm0.035$\\
           & Ly$\gamma/$Ly$\alpha$& 0.025 & 0.048 & ---\\
	   & Ly$\delta/$Ly$\alpha$& 0.010 & 0.023 & $\lsim0.063\pm0.013$\\
\enddata
\tablenotetext{a}{Evaluated at \ce\ temperature which maximizes ion fraction.}
\tablenotetext{b}{Evaluated at $kT_{\mathrm{e}}=4$~eV, although the 
temperature dependence is weak.}
\tablenotetext{c}{Error bars are 1-$\sigma$.}
\label{tab:Hratios}
\end{deluxetable}

\newpage
\begin{deluxetable}{lccc}
\tablecolumns{4}
\tablewidth{3.2in}
\tablehead{\colhead{Ion} & \colhead{Line Ratio} &\colhead{\rec \tablenotemark{a}}& \colhead{Observed\tablenotemark{b}}}
\startdata
\ion{N}{6} & He$\beta/f$ & 0.047 & $\lsim0.12\pm0.06$\\
           & He$\gamma/f$& 0.016 & $0.091\pm0.017$\\
           & He$\delta/f$& 0.008 & $0.038\pm0.017$\\
\ion{O}{7} & He$\beta/f$ & 0.051 & $0.138\pm0.019$\\
           & He$\gamma/f$& 0.017 & $0.083\pm0.012$\\
	   & He$\delta/f$& 0.008 & $0.065\pm0.010$\\
\enddata
\tablenotetext{a}{Evaluated at $kT_{\mathrm{e}}=3$~eV for \ion{N}{6} and 4~eV for \ion{O}{7}.}
\tablenotetext{b}{Error bars are 1-$\sigma$.}
\label{tab:Heratios}
\end{deluxetable}

\newpage
\begin{deluxetable}{lcccc}
\tablecolumns{5}
\tablewidth{4.1in}
\tablehead{\colhead{Ion} & \colhead{Ratio} & \colhead{\ce \tablenotemark{a}} & \colhead{\rec \tablenotemark{b}} & \colhead{Observed\tablenotemark{c}}}
\startdata
\ion{N}{6}	& $R=f/i$	&5.0--5.5&6.0--7.0&$6.9\pm1.1$\\
\ion{O}{7}	& $R=f/i$	&4.0--4.3&3.4--4.0&$5.5\pm0.8$\\
\ion{N}{6}	& $G=(f+i)/r$	&0.5--1.0&4.0--5.5&$2.5\pm0.2$\\
\ion{O}{7}  	& $G=(f+i)/r$	&0.5--1.1&4.0--5.3&$2.5\pm0.2$\\
\enddata
\tablenotetext{a}{Valid over realistic range of collisional $kT_{\mathrm{e}}$ for each ion \cite{smith}.}
\tablenotetext{b}{Valid for $kT_{\mathrm{e}}=0.1$--100~eV \cite{model}.}
\tablenotetext{c}{Error bars are 1-$\sigma$.}
\label{tab:rif}
\end{deluxetable}

\newpage
\begin{deluxetable}{lccc}
\tablewidth{3.8in}
\tablecolumns{4}
\tablehead{\colhead{Ion} & 
\colhead{$N^{\mathrm{rad}}_{\mathrm{ion}}$ [cm$^{-2}$]}  & \colhead{$kT_{\mathrm{e}}$ [eV]\tablenotemark{a}} & EM [10$^{64}$~cm$^{-3}$]\tablenotemark{b}}
\startdata
\ion{C}{5}	&8E17	&2.5	& 6.7\\
\ion{C}{6}	&9E17	&4.0	& 1.8\\
\ion{N}{6}	&6E17	&3.0	& 6.5\\
\ion{N}{7}	&6E17	&4.0	& 1.7\\
\ion{O}{7}	&1.1E18	&4.0	& 1.5\\
\ion{O}{8}	&1E18	&4.0	& 0.32\\
\ion{Ne}{9}	&3E17	&{\em4} & {\em 0.42}\\
\ion{Ne}{10}	&2.5E17	&{\em4} & {\em 0.097}\\
\ion{Mg}{11}	&2E17	&{\em4} & {\em 0.31}\\
\ion{Mg}{12}	&2E17	&{\em4} & {\em 0.092}\\
\ion{Si}{13}	&2E17	&{\em4} & {\em 0.13}\\
\ion{Si}{14}	&2E17	&{\em4} & {\em 0.042}\\
\enddata
\tablenotetext{a}{$kT_{\mathrm{e}}$ values in italics are arbitrarily assumed.}
\tablenotetext{b}{Assumes Solar abundance and fractional ionic abundance 
$f_{i+1}=0.5$.  Values in italics correspond to italicized (and arbitrary) $kT_{\mathrm{e}}$ values.}
\label{tab:finalfit}
\end{deluxetable}

\end{document}